\documentclass[aps,prd,final,twocolumn,letterpaper,superscriptaddress,nofootinbib,notitlepage]{revtex4-1}

\usepackage{graphicx}                       
\usepackage[outdir=./]{epstopdf}
\usepackage{amsmath}
\usepackage{amssymb}
\usepackage{amsthm}
\usepackage{dsfont}       
\usepackage[usenames,dvipsnames,table]{xcolor}
\usepackage[format=plain, font=small, labelfont=bf, justification = raggedright]{caption}
\usepackage{array}  
\usepackage{overpic}
\usepackage{placeins} 
\usepackage{fancyhdr}
\usepackage{colortbl}
\usepackage{multirow} 
\usepackage{hyperref} 
\hypersetup{colorlinks = true,linkcolor = blue, breaklinks = true}

\newcommand{\eq}[1]{(\ref{#1})}
\newcommand{\Eq}[1]{Eq.~\eq{#1}}
\newcommand{\Eqs}[1]{Eqs.~\eq{#1}}

\newcommand{\Sec}[1]{Sec.~\ref{#1}}

\renewcommand{\Ref}[1]{Ref.~\cite{#1}}

\newcommand{\Refs}[1]{Refs.~\cite{#1}}
\newcommand{\App}[1]{Appendix~\ref{#1}}

\newcommand{\ie}{{i.e., }}

\newcommand{\msf}[1]{\mathsf{#1}}

\newcommand{\bra}[1]{\langle#1 |}
\newcommand{\ket}[1]{|#1 \rangle}
\newcommand{\braket}[2]{\langle#1 |  #2 \rangle}
\newcommand{\com}[2]{\left[#1, #2\right]}

\newcommand{\oper}[1]{\smash{\hat{#1}}}

\newcommand{\fourier}[1]{\smash{\widetilde{#1}}}

\newcommand{\dd}{\mathrm{d}}

\newcommand{\Vect}[1]{{\boldsymbol{\rm #1}}}
\newcommand{\VectOp}[1]{\oper{\Vect{#1}}}

\newcommand{\Mat}[1]{\msf{#1}}
\newcommand{\IMat}[1]{\Mat{I}_{#1}}
\newcommand{\OMat}[1]{\Mat{0}_{#1}}
\newcommand{\dubdot}{\text{\large :}}

\newcommand{\IdentOp}{\oper{\mathds{1}}}

\newcommand{\nullFrac}{\vphantom{\frac{}{}}}

\newcommand{\Stroke}[1]{\text{\ooalign{ $#1$\cr \hidewidth\raise.225ex \hbox{$-\mkern.5mu$}\cr}}}

\begin{document}
\setlength{\parskip}{0pt}
\setlength{\belowcaptionskip}{0pt}


\title{Exactly unitary discrete representations of the metaplectic transform for linear-time algorithms}
\author{N. A. Lopez}
\affiliation{Department of Astrophysical Sciences, Princeton University, Princeton, New Jersey 08544, USA}
\author{I. Y. Dodin}
\affiliation{Department of Astrophysical Sciences, Princeton University, Princeton, New Jersey 08544, USA}
\affiliation{Princeton Plasma Physics Laboratory, Princeton, New Jersey 08543, USA}

\begin{abstract}

The metaplectic transform (MT), a generalization of the Fourier transform sometimes called the linear canonical transform, is a tool used ubiquitously in modern optics, for example, when calculating the transformations of light beams in paraxial optical systems. The MT is also an essential ingredient of the geometrical-optics modeling of caustics that was recently proposed by the authors. In particular, this application relies on the near-identity MT (NIMT); however, the NIMT approximation used so far is not exactly unitary and leads to numerical instability. Here, we develop a discrete MT that is exactly unitary, and approximate it to obtain a discrete NIMT that is also unitary and can be computed in linear time. We prove that the discrete NIMT converges to the discrete MT when iterated, thereby allowing the NIMT to compute MTs that are not necessarily near-identity. We then demonstrate the new algorithms with a series of examples.

\end{abstract}

\maketitle

\pagestyle{fancy}
\lhead{Lopez \& Dodin}
\rhead{Unitary NIMT}
\thispagestyle{empty}


\section{Introduction}
\label{sec:intro}

It is often of interest in optics to compute the metaplectic transform (MT), also called the linear canonical transform, of a given wavefield $\psi$~\cite{Healy16}. For example, just as the action of a lens is to transform $\psi$ to its Fourier transform at the focal plane, the action of a general paraxial optical system is to transform $\psi$ to its MT at the output plane~\cite{Collins70,Bacry81,Simon00,Wolf18}. As another example, the MT and the near-identity MT (NIMT) can be used for calculating wave caustics within a reduced framework called metaplectic geometrical optics, which avoids the usual singularities~\cite{Lopez20a,Lopez21a}. Hence, the ability to calculate the MT and NIMT accurately and efficiently is important for practical applications.

In recent years, many fast algorithms have been developed to compute the MT for one-dimensional (1-D) and 2-D fields~\cite{Ozaktas96,Hennelly05b,Healy10,Koc10a,Ding12,Pei16,Sun18a,Healy18}. These algorithms often involve the fast Fourier transform, since the MT has a simple spectral representation; as such, they often scale as $O(N \log_2 N)$, where $N$ is the number of sample points. Comparatively less developed are fast algorithms to compute the NIMT. Reference~\cite{Lopez19a} showed that algorithms designed specifically for the NIMT can be computed with $O(N)$ complexity. This is potentially a significant speedup compared to other MT algorithms, but unfortunately, the fast NIMT of \Ref{Lopez19a} suffered from numerical instability, which can be traced back to a loss of unitarity.

Here, we improve upon this work and develop a fast algorithm that calculates an exactly unitary approximation of the NIMT. To do this, we first develop a discrete matrix representation of the MT that is exactly unitary. Through use of a diagonal Pad\'e approximation, we then obtain a discrete NIMT that is also exactly unitary. This discrete NIMT can be used `as is', or as we prove, it can also be iterated to perform arbitrary MTs without experiencing numerical instability. We show that our fast NIMT can still be computed in time that scales linearly with the number of grid points. We therefore expect this algorithm to be useful in a broad range of applications. We should caution, however, that although $K \gg 1$ iterated NIMTs can perform the FT in $O(KN)$ computations in principle, this scaling is less appealing than the $O(N \log_2 N)$ scaling of the FFT at practical $K$ and $N$, so this application is not put forth here.

This paper is organized as follows. In \Sec{sec:operMT}, we review the pseudo-differential representation of the MT that was originally developed in \Ref{Lopez19a}. In \Sec{sec:dMT}, we use the pseudo-differential representation of the MT to develop a discrete representation that is exactly unitary. We also briefly comment on how the discrete MT can be approximately computed in linear time. In \Sec{sec:dNIMT}, we develop a discrete NIMT from the discrete MT, then show that the discrete NIMT is exactly unitary and can be computed in linear time. In \Sec{sec:iterNIMT}, we show how the new NIMT algorithm can be iterated to perform an arbitrary MT with robust local and global convergence. In \Sec{sec:example}, we demonstrate the new algorithms in a series of examples. In \Sec{sec:concl}, we summarize our main conclusions. Auxiliary calculations are presented in the appendix.

\section{Operator representation of the metaplectic transform}
\label{sec:operMT}

\subsection{Definition of the MT}

Consider the propagation of a wavefield through an $(m+1)$-D paraxial optical setup, where $m$ is the number of dimensions transverse to the optical axis. Let $\Vect{q}$ be coordinates on the plane transverse to the optical axis, and let $\Vect{p}$ be coordinates on the corresponding $m$-D space of wavevectors that are Fourier-dual to $\Vect{q}$. Collectively, $(\Vect{q}, \Vect{p})$ define coordinates for the $2m$-D phase space that is populated by the rays of geometrical optics. A general paraxial setup can be represented through its linear action on this phase space by a $2m\times 2m$ matrix $\Mat{S}$ as
\begin{equation}
    \begin{pmatrix}
        \Vect{Q} \\
        \Vect{P}
    \end{pmatrix}
    =
    \Mat{S}
    \begin{pmatrix}
        \Vect{q} \\
        \Vect{p}
    \end{pmatrix}
    ,
    \quad
    \Mat{S}
    =
    \begin{pmatrix}
        \Mat{A} & \Mat{B} \\
        \Mat{C} & \Mat{D}
    \end{pmatrix}
    ,
\end{equation}

\noindent where $(\Vect{Q}, \Vect{P})$ are coordinates and wavevectors on the output plane, and the matrices $\Mat{A}$, $\Mat{B}$, $\Mat{C}$, and $\Mat{D}$ are all size $m \times m$. (In this context, $\Mat{S}$ is commonly referred to as the $ABCD$ ray transfer matrix~\cite{Kogelnik66}.) When the paraxial setup is lossless, then $\Mat{S}$ is symplectic, satisfying
\begin{equation}
    \Mat{S}
    \begin{pmatrix}
        \OMat{m} & \IMat{m}\\
        -\IMat{m} & \OMat{m}
    \end{pmatrix} \Mat{S}^\intercal 
    = 
    \begin{pmatrix}
        \OMat{m} & \IMat{m}\\
        -\IMat{m} & \OMat{m}
    \end{pmatrix}
    ,
    \label{eq:sympS}
\end{equation}

\noindent where $\OMat{m}$ and $\IMat{m}$ are the $m \times m$ null and identity matrices respectively. Equation \eq{eq:sympS} also implies the relations~\cite{Luneburg64}
\begin{subequations}
    \begin{align}
        \label{eq:symplec1}
        \Mat{A} \Mat{D}^\intercal - \Mat{B} \Mat{C}^\intercal &= \IMat{m} \, ,\\
        \label{eq:symplec2}
        \Mat{A}^\intercal \Mat{D} - \Mat{C}^\intercal \Mat{B} &= \IMat{m} \, ,\\
        \label{eq:symplec3}
        \Mat{A} \Mat{B}^\intercal - \Mat{B} \Mat{A}^\intercal &= \OMat{m} \, ,\\
        \label{eq:symplec4}
        \Mat{B}^\intercal \Mat{D} - \Mat{D}^\intercal \Mat{B} &= \OMat{m} \, ,\\
        \label{eq:symplec5}
        \Mat{C}^\intercal \Mat{A} - \Mat{A}^\intercal \Mat{C} &= \OMat{m} \, ,\\
        \label{eq:symplec6}
        \Mat{D} \Mat{C}^\intercal - \Mat{C} \Mat{D}^\intercal &= \OMat{m} \, ,
    \end{align}
\end{subequations}

\noindent where $^\intercal$ denotes matrix transpose.

Suppose that the transverse field profiles at the input and output planes are given by $\psi(\Vect{q})$ and $\Psi(\Vect{Q})$, respectively. Then in the paraxial limit, $\Psi(\Vect{Q})$ is obtained from $\psi(\Vect{q})$ by the MT corresponding to $\Mat{S}$, given either by the integral expression~\cite{Collins70,Moshinsky71,Littlejohn86a}
\begin{align}
    \Psi(\Vect{Q}) 
    &= \pm
    \frac{ 
        \exp\left(
            \frac{i}{2}\Vect{Q}^\intercal \Mat{D} \Mat{B}^{-1} \Vect{Q}
        \right)
    }{
        (2\pi i)^{\frac{m}{2}}
        \sqrt{\det{\Mat{B}}}
    }
    \nonumber\\
    &\hspace{0mm} \times 
    \int \dd\Vect{q} \,
    \exp\left(
        \frac{i}{2}\Vect{q}^\intercal \Mat{B}^{-1}\Mat{A} \Vect{q} - i\Vect{q}^\intercal \Mat{B}^{-1} \Vect{Q}
    \right)
    \psi(\Vect{q})
    ,
    \label{eq:intMT}
\end{align}

\noindent or equivalently, by the pseudo-differential expression~\cite{Lopez19a}
\begin{align}
    \Psi(\Vect{Q}) 
    &= \pm
    \frac{
        \exp\left(
            \frac{i}{2} \Vect{q}^\intercal \Mat{A}^\intercal \Mat{C} \Vect{q}
        \right)
    }{
        \sqrt{\det \Mat{A} }
    }
    \nonumber\\
    &\hspace{0mm}\times
    \left.
        \exp\left(
            \frac{i}{2} \, \Mat{A}^{-1} \Mat{B} \dubdot \nabla \nabla 
        \right) \psi(\Vect{q})
    \right|_{\Vect{q} = \Mat{A}^{-1}\Vect{Q}} 
    ,
    \label{eq:PMT}
\end{align} 

\noindent where $\nabla \equiv \dd/ \dd \Vect{q}$. Note that \Eq{eq:intMT} requires $\det \Mat{B} \neq 0$, while \Eq{eq:PMT} requires $\det \Mat{A} \neq 0$. For our purposes in developing truncations of the MT for when $\Mat{S}$ is near-identity, this latter requirement will always hold. 

Note that the MT is a unitary transformation from $\psi(\Vect{q})$ to $\Psi(\Vect{Q})$, which is consistent with the condition that the paraxial setup be lossless. Also note that the MT satisfies the composition property
\begin{equation}
    \oper{M}(\Mat{S}_1)\oper{M}(\Mat{S}_2) = \oper{M}(\Mat{S}_1\Mat{S}_2)
    ,
    \label{eq:MTcomp}
\end{equation}

\noindent which allows the transformation of $\psi$ through a cascaded optical setup to be computed with ease.

\subsection{Manifestly unitary form of the MT}

It will be useful in the forthcoming sections to recast \Eq{eq:PMT} into a manifestly unitary form. To do so, let us introduce the abstract operator formalism of quantum mechanics~\cite{Stoler81}. In this formalism, $\psi(\Vect{q})$ and $\Psi(\Vect{Q})$ are represented by state vectors $\ket{\psi}$ and $\ket{\Psi}$, which are related by the linear relation
\begin{equation}
    \ket{\Psi} = \oper{M}(\Mat{S}) \ket{\psi}
    ,
    \label{eq:ketMT}
\end{equation}

\noindent where $\oper{M}(\Mat{S})$ is the abstract operator that enacts the MT \eq{eq:PMT}. We also introduce the Hermitian operator $\VectOp{q}$, which enacts multiplication by $\Vect{q}$, and the Hermitian operator $\VectOp{p}$, which enacts differentiation by $\Vect{q}$. The operators $\VectOp{q}$ and $\VectOp{p}$ satisfy the canonical commutation relation
\begin{equation}
    \com{
        \oper{q}_j
    }{
        \oper{p}_k
    }
    =
    i \delta_{jk} \IdentOp
    ,
    \quad j,k = 1, \ldots, m
    ,
    \label{eq:qpCOM}
\end{equation}

\noindent where $\delta_{jk}$ is the Kronecker delta and $\IdentOp$ is the identity operator.

Since $\VectOp{q}$ is Hermitian, it possesses a complete set of eigenvectors; the eigenvector corresponding to eigenvalue $\Vect{q}$ is denoted by $\ket{\Vect{q}}$ and satisfies
\begin{equation}
    \VectOp{q} \ket{\Vect{q}}
    =
    \Vect{q} \ket{\Vect{q}}
    .
\end{equation}

\noindent The wavefields $\psi(\Vect{q})$ and $\Psi(\Vect{Q})$ are obtained from $\ket{\psi}$ and $\ket{\Psi}$ by projection onto the basis $\{ \ket{\Vect{q}}\}$ as
\begin{equation}
    \psi(\Vect{q}) \doteq \braket{\Vect{q}}{\psi}
    , \quad
    \Psi(\Vect{Q}) \doteq \braket{\Vect{Q}}{\Psi}
    .
\end{equation}

\noindent (Note that $\ket{\Vect{Q}}$ is an eigenvector of $\VectOp{q}$ with eigenvalue $\Vect{Q}$.) Also, for any function $f$,
\begin{subequations}
    \begin{align}
        f(\Vect{q}) \psi(\Vect{q}) 
        &= \bra{\Vect{q}} f(\VectOp{q}) \ket{\psi}
        , \\
        f(\nabla) \psi(\Vect{q}) 
        &= \bra{\Vect{q}} f(i \VectOp{p}) \ket{\psi}
        .
    \end{align}
\end{subequations}%

Let us also introduce the inverse dilation operator $\oper{D}_{\Mat{A}}$ that acts on the state vectors as
\begin{equation}
    \bra{\Vect{q} }\oper{D}_{\Mat{A}} \ket{\psi} \doteq 
    \sqrt{\det \Mat{A}^{-1} } \, \psi(\Mat{A}^{-1}\Vect{q})
    ,
\end{equation}

\noindent where the factor $\sqrt{\det \Mat{A}^{-1} }$ ensures that $\oper{D}_{\Mat{A}}$ is unitary. Then, by comparing \Eq{eq:PMT} with \Eq{eq:ketMT}, one sees that $\oper{M}(\Mat{S})$ takes the form~\cite{Lopez19a}
\begin{equation}
    \oper{M}(\Mat{S})
    =
    \pm
    \oper{D}_{\Mat{A}}
    \exp\left(
        \frac{i}{2} \VectOp{q}^\intercal \Mat{A}^\intercal \Mat{C} \VectOp{q}
    \right)
    \exp\left(
        - \frac{i}{2} \VectOp{p}^\intercal \Mat{A}^{-1} \Mat{B} \VectOp{p}
    \right)
    .
    \label{eq:operMTold}
\end{equation}

\noindent Notice that $\oper{D}_\Mat{A}$ can also be written as (\App{app:verify})
\begin{equation}
    \oper{D}_{\Mat{A}}
    =
    \exp\left[ 
        i \frac{
            \VectOp{q}^\intercal \left(\log \Mat{A}^{-\intercal}\right) \VectOp{p} + \VectOp{p}^\intercal \left(\log \Mat{A}^{-1}\right) \VectOp{q}
        }{2}
    \right] 
    ,
    \label{eq:dilOPER}
\end{equation}

\noindent which is a multidimensional analog of the familiar quantum squeezing operator~\cite{Scully12}. (Here, $^{-\intercal}$ denotes the inverse-matrix transpose.) Hence, \Eq{eq:operMTold} can be written in the manifestly unitary form
\begin{align}
    \oper{M}(\Mat{S}) 
    &= \pm 
    \exp\left[ 
        i \frac{
            \VectOp{q}^\intercal \left(\log \Mat{A}^{-\intercal}\right) \VectOp{p} 
            + \VectOp{p}^\intercal \left(\log \Mat{A}^{-1}\right) \VectOp{q}
        }{2}
    \right] 
    \nonumber\\
    &\hspace{0mm}\times
    \exp\left(
        \frac{i}{2} \VectOp{q}^\intercal \Mat{A}^\intercal \Mat{C} \VectOp{q}
    \right)
    \, 
    \exp\left(
        - \frac{i}{2} \VectOp{p}^\intercal \Mat{A}^{-1} \Mat{B} \VectOp{p}
    \right)
    .
    \label{eq:operMT}
\end{align}

The three operator exponentials that enter \Eq{eq:operMT} can be understood as the individual MTs induced by the following block matrices:
\begin{subequations}
    \begin{align}
        \Mat{G} 
        &\doteq
        \begin{pmatrix}
            \Mat{A} & \OMat{m}\\
            \OMat{m} & \Mat{A}^{-\intercal}
        \end{pmatrix}
        , \\
        \Mat{L}
        &\doteq
        \begin{pmatrix}
            \IMat{m} & \OMat{m} \\
            \Mat{A}^\intercal \Mat{C} & \IMat{m}
        \end{pmatrix}
        , \\
        \Mat{U}
        &\doteq
        \begin{pmatrix}
            \IMat{m} & \Mat{A}^{-1}\Mat{B} \\
            \OMat{m} & \IMat{m}
        \end{pmatrix}
        .
    \end{align}
\end{subequations}

\noindent In this sense, \Eq{eq:operMT} is the decomposition of the MT induced by the following decomposition of the corresponding symplectic matrix $\Mat{S}$:
\begin{equation}
    \Mat{S}
    = \Mat{G} \Mat{L} \Mat{U}
    ,
    \label{eq:decompS}
\end{equation}

\noindent which can be viewed as a modified LDU decomposition. By recognizing $\Mat{G}$, $\Mat{L}$, and $\Mat{U}$ as ray-transfer matrices~\cite{Kogelnik66}, a physical interpretation to \Eq{eq:operMT} is readily obtained: $\oper{M}(\Mat{S})$ represents the action of a paraxial optical system consisting of propagation in uniform media $\Mat{U}$ (generally anisotropic), followed by a thin lens $\Mat{L}$ (generally asymmetric), followed by magnification $\Mat{G}$.


\section{Discrete metaplectic transform}
\label{sec:dMT}

\subsection{Derivation}

In the form \eq{eq:operMT}, the MT can be readily discretized. Let us consider the 1-D case ($m = 1$) for simplicity. In this case, \Eq{eq:operMT} has the $q$-space representation
\begin{align}
    \oper{M}(\Mat{S}) &= \pm 
    \exp\left[ 
        \frac{\log A^{-1} }{2}
        \left(
             q \frac{\dd}{\dd q} 
             + \frac{\dd}{\dd q} q
        \right)
    \right] 
    \nonumber\\
    &\times
    \exp\left(
        \frac{i}{2} AC q^2
    \right)
    \, 
    \exp\left(
        \frac{i B }{2A } 
        \frac{\dd^2}{\dd q^2}
    \right)
    .
    \label{eq:1DMT}
\end{align}

\noindent Let us consider an $N$-point discretization of $q$-space given by the set $\{ q_j \}$, $j = 1, \ldots, N$. We assume that the set $\{ q_j \}$ is distinct and lexicographically ordered such that $q_j < q_k$ when $j < k$. Then, functions of $q$ are discretized by the values on $\{ q_j \}$, and are represented by vectors of length $N$ as
\begin{equation}
    \psi(q)
    \mapsto
    \Vect{\psi} 
    \doteq
    \begin{pmatrix}
        \psi_1 &
        \ldots &
        \psi_N
    \end{pmatrix}^\intercal
    , \quad
    \psi_j \doteq \psi(q_j)
    .
\end{equation}

Similarly, operators are represented as $N \times N$ matrices. In particular, the coordinate operator $q$ is represented by the diagonal matrix
\begin{equation}
    \Mat{q} \doteq
    \begin{pmatrix}
        q_1 & & \\
        & \ddots & \\
        & & q_N
    \end{pmatrix}
    ,
\end{equation}

\noindent and analytic functions of $q$ are discretized by the formal replacement $q \mapsto \Mat{q}$. Pseudo-differential operators are represented by functions of finite-difference matrices with certain restrictions. For example, to discretize analytic functions of $\dd/\dd q$, we introduce a family of $N \times N$ finite-difference matrices on $\{ q_i\}$, denoted by the set $\{\delta_\ell \}$ with $\ell \ge 1$ such that $\delta_{\ell} \, \Vect{\psi}$ is a suitable discretization of $\dd^\ell \psi /\dd q^\ell$. We then require $\delta_1$ to be the family generator:
\begin{equation}
    \delta_{\ell} = \delta_1^\ell
    , \quad
    \ell = 1, 2, \ldots
    ,
    \label{eq:deltaREQ}
\end{equation}

\noindent and also to be skew-Hermitian, so that $\delta_1$ faithfully mimics the purely imaginary eigenspectrum of $\dd/\dd q$. Then, any analytic function of $\dd/ \dd q$ is discretized by the formal replacement $\dd/ \dd q \mapsto \delta_1$.

To discretize analytic functions of $\dd^2 / \dd q^2$, we can simply perform the formal replacement $\dd^2 / \dd q^2 \mapsto \delta_1^2$. However, it is often more convenient to introduce an additional family of $N \times N$ even-order finite-difference matrices on $\{ q_i \}$, denoted by the set $\{\Delta_{2k}\}$ with $k \ge 1$ such that $\Delta_{2k} \, \Vect{\psi}$ is a suitable discretization of $\dd^{2k} \psi /\dd q^{2k}$. We then require $\Delta_2$ to be the family generator:
\begin{equation}
    \Delta_{2k} = \Delta_2^k
    , \quad
    k = 1, 2, \ldots 
    ,
    \label{eq:DeltaREQ}
\end{equation}

\noindent and also to be Hermitian and negative semi-definite, so that $\Delta_2$ faithfully mimics the negative semi-definite eigenspectrum of $\dd^2 / \dd q^2$. (Note that $\{\Delta_{2k} \}$ need not coincide with $\delta_{\ell}$ at even $\ell$, but clearly, any suitable $\{ \delta_\ell \}$ also constitutes a suitable $\{ \Delta_{2k} \}$.) Then, any analytic function of $\dd^2/ \dd q^2$ is discretized by the formal replacement $\dd^2 / \dd q^2 \mapsto \Delta_2$. In particular, \Eq{eq:1DMT} is discretized to yield the discrete MT (dMT), given as
\begin{align}
    \oper{M}(\Mat{S})
    \mapsto
    \Mat{M}(\Mat{S})
    &\doteq
    \pm 
    \exp\left(
        \log A^{-1} \,
        \frac{
            \Mat{q} \, \delta_1
            + \delta_1 \Mat{q}
        }{2}
    \right)
    \nonumber\\
    &\times
    \exp
    \left(
        \frac{i A C}{2} \Mat{q}^2
    \right)
    \,
    \exp\left(
        \frac{i B}{2A} 
        \Delta_2
    \right)
    .
    \label{eq:dMT}
\end{align}

When $\{ q_j \}$ are equally spaced with grid spacing $h$, then suitable choices of $\delta_1$ and $\Delta_2$ are the central-difference matrices. These matrices are of Toeplitz form~\cite{Strang14}; explicitly, a 2nd-order finite-difference scheme has %
\begin{subequations}
    \label{eq:2ndDELTA}
    \begin{align}
        \delta_1^{(2)}
        &= \frac{1}{2h}
        \begin{pmatrix}
             0 &      1 &    \\
            -1 & \ddots &  1 \\
               &     -1 &  0 \\
        \end{pmatrix}
        ,
        \\
        \Delta_2^{(2)}
        &=
        \frac{1}{h^2}
        \begin{pmatrix}
            -2 &      1 &    \\
             1 & \ddots &  1 \\ 
               &      1 & -2 \\
        \end{pmatrix}
        ,
    \end{align}
\end{subequations}

\noindent while a 4th-order finite-difference scheme has%
\begin{subequations}
    \label{eq:4thDELTA}
    \begin{align}
        \delta_1^{(4)}
        &= \frac{1}{12h}
        \begin{pmatrix}
             0 &      8 &     -1 &    \\
            -8 & \ddots & \ddots & -1 \\
             1 & \ddots & \ddots &  8 \\
               &      1 &     -8 &  0 \\
        \end{pmatrix}
        , \\
        \Delta_2^{(4)}
        &=
        \frac{1}{12h^2}
        \begin{pmatrix}
            -30 &     16 &     -1 &     \\
             16 & \ddots & \ddots &  -1 \\ 
             -1 & \ddots & \ddots &  16 \\
                &     -1 &     16 & -30 \\
        \end{pmatrix}
        ,
    \end{align}
\end{subequations}

\noindent and a 6th-order finite-difference scheme has%
\begin{subequations}
    \label{eq:6thDELTA}
    \begin{align}
        \delta_1^{(6)}
        &= \frac{1}{60h}
        \begin{pmatrix}
              0 &     45 &     -9 &      1 &    \\
            -45 & \ddots & \ddots & \ddots &  1 \\
              9 & \ddots & \ddots & \ddots & -9 \\
             -1 & \ddots & \ddots & \ddots & 45 \\
                &     -1 &      9 &    -45 &  0 \\
        \end{pmatrix}
        , \\
        \Delta_2^{(6)}
        &=
        \frac{1}{180h^2}
        \begin{pmatrix}
            -490 &    270 &    -27 &      2 &      \\
             270 & \ddots & \ddots & \ddots &    2 \\ 
             -27 & \ddots & \ddots & \ddots &  -27 \\
               2 & \ddots & \ddots & \ddots &  270 \\
                 &      2 &    -27 &    270 & -490 \\
        \end{pmatrix}
        .
    \end{align}
\end{subequations}

\noindent Clearly, each $\delta_1$ is skew-symmetric while each $\Delta_2$ is symmetric, as required; hence, the dMT \eq{eq:dMT} is unitary, as we show explicitly below. (In contrast, forward and backward finite-difference matrices do not have the required symmetry for use in the dMT.) In \Sec{sec:example}, we shall compare the 2nd-order, 4th-order, and 6th-order dMTs that use \Eqs{eq:2ndDELTA}, \eq{eq:4thDELTA}, and \eq{eq:6thDELTA}, respectively.


\subsection{Unitarity verification}
\label{sec:dMTunit}

Since $\Delta_2$ is Hermitian, the matrix $iB \Delta_2/2A $ is skew-Hermitian, which implies that $\exp(iB \Delta_2/2A)$ is unitary. Likewise, the matrix $\exp(i A C \Mat{q}^2/2)$ is unitary as well. Lastly, since $\Mat{q}$ is Hermitian and $\delta_1$ is skew-Hermitian, the matrix $\Mat{q} \, \delta_1+ \delta_1\Mat{q}$ is skew-Hermitian:
\begin{align}
    ( \Mat{q} \, \delta_1+ \delta_1\Mat{q} )^\dagger
    = \delta_1^\dagger \Mat{q} + \Mat{q} \delta_1^\dagger
    = - \left( \Mat{q} \, \delta_1+ \delta_1\Mat{q} \right)
    .
\end{align}

\noindent Consequently, $\exp[- \log{A} ( \Mat{q} \, \delta_1+ \delta_1\Mat{q} )/2 ]$ is unitary. As the product of unitary matrices, we conclude that $\Mat{M}(\Mat{S})$ is unitary as well.


\subsection{Runtime estimate}
\label{sec:dMTtime}

Here we estimate the computational complexity of the dMT for a uniform grid. First, let us estimate the cost to construct the matrix factors in $\Mat{M}(\Mat{S})$. Although formally dense, the matrix exponential of a banded matrix is `pseudo-sparse' and essentially banded, since the matrix elements rapidly decrease away from the main diagonal~\cite{Iserles00}. This means that the matrix exponential can be computed (approximately) in linear time, \ie $O(N)$\cite{Benzi07}. Thus, we estimate the construction of the three matrix factors in $\Mat{M}(\Mat{S})$ to be $O(N)$. Alternatively, \Eqs{eq:exp1} and \eq{eq:exp2} below can be directly approximated in $O(N)$ time by using a low-order `scaling-squaring' method based on Pad\'e~\cite{AlMohy09} or truncated Taylor approximants~\cite{AlMohy11}.

Assuming the matrix factors have been constructed, we next estimate the cost to perform the dMT as
\begin{equation}
    \Vect{\Psi} = \Mat{M}(\Mat{S}) \Vect{\psi}
    .
\end{equation}

\noindent Suppose that the matrix exponentials involving $\Delta_2$ and $\delta_1$ have been approximated by banded matrices with bandwidths of $b_\Delta$ and $b_\delta$ respectively. Since matrix-vector multiplication involving a banded matrix of bandwidth $b$ requires $O(bN)$ operations, computing
\begin{equation}
    \Vect{v}_1 \doteq 
    \exp
    \left(
        \frac{i B}{2 A}
        \Delta_2
    \right)
    \Vect{\psi}
    \label{eq:exp1}
\end{equation}

\noindent requires $O(b_\Delta N)$ operations. Similarly, computing
\begin{equation}
    \Vect{v}_2 \doteq 
    \exp
    \left(
        \frac{i A C}{2}
        \Mat{q}^2
    \right)
    \Vect{v}_1
\end{equation}

\noindent requires $O(N)$ operations, and computing 
\begin{equation}
    \Vect{\Psi}
    =
    \exp\left(
        \log A^{-1} \,
        \frac{
            \Mat{q} \, \delta_1
            + \delta_1 \Mat{q}
        }{2}
    \right)
    \Vect{v}_2
    \label{eq:exp2}
\end{equation}

\noindent requires $O(b_\delta N)$ operations. Since these operations are performed in sequence, we conclude that the dMT can be approximately computed in linear time. (Note that the approximation is due to representing the matrix exponentials with banded matrices.)


\section{Discrete near-identity metaplectic transform}
\label{sec:dNIMT}

\subsection{Derivation}

Let us now consider simplifications to \Eq{eq:dMT} when $\Mat{S}$ is near-identity, that is, $|A| \sim 1$ and $|B| \sim |C| \ll 1$. In \Ref{Lopez19a}, this limit was considered by performing a Taylor expansion of the matrix exponential; however, the NIMT that resulted was not unitary and suffered from numerical instability as a consequence. To remedy this, here we shall instead use a [1/1] Pad\'e approximation~\cite{Press07,Olver10a} for the matrix exponential, given explicitly as
\begin{equation}
    \exp(\Mat{H})
    \approx 
    \left(
        \IMat{N} 
        - \frac{\Mat{H}}{2}
    \right)
    \left(
        \IMat{N} 
        + \frac{\Mat{H}}{2}
    \right)
    , \quad 
    \| \Mat{H} \| \ll 1
    .
    \label{eq:11pade}
\end{equation}

\noindent Hence, we approximate $\Mat{M}(\Mat{S}) \approx \Mat{N}(\Mat{S})$, where we have introduced
\begin{align}
    \Mat{N}(\Mat{S}) 
    =
    &\left(
        \IMat{N}
        + \log A
        \frac{
            \Mat{q} \, \delta_1
            + \delta_1 \Mat{q}
        }{4}
    \right)^{-1}
    \nonumber\\
    &\times
    \left(
        \IMat{N}
        - \log A
        \frac{
            \Mat{q} \, \delta_1
            + \delta_1 \Mat{q}
        }{4}
    \right)
    \exp
    \left(
        \frac{i A C}{2} \Mat{q}^2
    \right)
    \nonumber\\
    &\times
    \left(
        \IMat{N}
        - \frac{i B}{4A} \Delta_2
    \right)^{-1}
    \left(
        \IMat{N}
        + \frac{i B}{4A} \Delta_2
    \right)
    ,
    \label{eq:dNIMT}
\end{align}

\noindent and we have chosen the overall $+$ sign in \Eq{eq:operMT}. We call \Eq{eq:dNIMT} the discrete NIMT (dNIMT). Note that there is no need to approximate $\exp(i A C \Mat{q}^2/2)$ because it is diagonal and therefore trivial to compute. Also note that in fact, any diagonal ([$r/r$] with integer $r$) Pad\'e approximation is suitable for use in the dNIMT (as we show below); we choose the [1/1] approximation for simplicity.

\subsection{Unitarity verification}

It is well-known that the [1/1] Pad\'e approximation preserves the unitarity of matrix exponentials by acting as a Cayley transform for the argument of the matrix exponential~\cite{Eves80,Diele98,Iserles01,Golub13,Zhang20,Fu20}. Here, we show this fact explicitly. Suppose that $\Mat{H}$ is a skew-Hermitian matrix. It therefore possesses a complete set of eigenvectors with imaginary eigenvalues, denoted by $\Vect{\lambda}_j$ and $i \lambda_j$ with $\lambda_j$ real. Consequently, the set $\{ \Vect{\lambda}_j \}$ also satisfy the eigenvalue relation
\begin{subequations}
    \begin{gather}
        \left(
            \IMat{N}
            - \frac{\Mat{H}}{2}
        \right)^{-1}
        \left(
            \IMat{N}
            + \frac{\Mat{H}}{2}
        \right)
        \Vect{\lambda}_j
        =
        e^{2 i \theta_j} \Vect{\lambda}_j
        , \\
        \theta_j 
        \doteq \textrm{arg} 
        \left(
            1 
            + i \frac{\lambda_j}{2} 
        \right)
        .
    \end{gather}
\end{subequations}

\noindent Hence, the [1/1] Pad\'e approximation of the matrix exponential \eq{eq:11pade} for skew-Hermitian $\Mat{H}$ is unitary. Since the Pad\'e approximations included in \Eq{eq:dNIMT} are all of this form, we conclude that the dNIMT is unitary as well. (Note that with minor modifications, the above proof also holds for a general diagonal Pad\'e approximation.)

\subsection{Runtime estimate}

Let us now present a runtime estimate. As before, we restrict the analysis to a uniform grid. We shall also consider only the lowest-order approximations for $\delta_1$ and $\Delta_2$ given by \Eqs{eq:2ndDELTA}; the analysis for higher-order approximations is analogous. 

Since $\Delta_2$ is tridiagonal, computing
\begin{equation}
    \Vect{v}_3
    \doteq
    \left(
        \IMat{N}
        + \frac{i B}{4A} \Delta_2
    \right) \Vect{\psi}
\end{equation}

\noindent requires $O(3N)$ operations. Next, note that $\Vect{v}_4$, defined as
\begin{equation}
    \Vect{v}_4
    \doteq
    \left(
        \IMat{N}
        - \frac{i B}{4A} \Delta_2
    \right)^{-1} \Vect{v}_3
    ,
\end{equation}

\noindent is the solution to the tridiagonal linear system
\begin{equation}
    \left(
        \IMat{N}
        - \frac{i B}{4A} \Delta_2
    \right)
    \Vect{v}_4
    =
    \Vect{v}_3
    ,
\end{equation}

\noindent which can be obtained in $O(N)$ computations using a tridiagonal Gaussian elimination algorithm~\cite{Press07}. Next, since $\Mat{q}$ is diagonal, computing
\begin{equation}
    \Vect{v}_5
    \doteq
    \exp \left(
        \frac{i A C}{2} \Mat{q}^2
    \right)
    \Vect{v}_4
\end{equation}

\noindent requires $O(N)$ operations. Next, since $\delta_1$ is tridiagonal, computing
\begin{equation}
    \Vect{v}_6
    \doteq
    \left(
        \IMat{N}
        - \log A
        \frac{
            \Mat{q} \, \delta_1
            + \delta_1 \Mat{q}
        }{4}
    \right)
    \Vect{v}_5
\end{equation}

\noindent requires $O(3N)$ operations. Lastly, rather than directly computing
\begin{equation}
    \Vect{\Psi}
    \doteq
    \left(
        \IMat{N}
        + \log A
        \frac{
            \Mat{q} \, \delta_1
            + \delta_1 \Mat{q}
        }{4}
    \right)^{-1}
    \Vect{v}_6
    ,
\end{equation}

\noindent we obtain $\Vect{\Psi}$ by solving the tridiagonal linear system
\begin{equation}
    \left(
        \IMat{N}
        + \log A
        \frac{
            \Mat{q} \, \delta_1
            + \delta_1 \Mat{q}
        }{4}
    \right)
    \Vect{\Psi}
    =
    \Vect{v}_6
    ,
\end{equation}

\noindent which requires $O(N)$ computations. Thus, by performing these computations in sequence, the dNIMT can be computed in linear time, \ie $O(N)$.

\section{Convergence of iterated discrete Near-identity metaplectic transform}
\label{sec:iterNIMT}

\subsection{Derivation}

A finite (not near-identity) MT corresponding to a desired symplectic transformation $\fourier{\Mat{S}}$ can be iteratively computed with the NIMT by finding a path in the space of symplectic matrices, denoted $\Mat{S}(t)$, such that $\Mat{S}(0) = \IMat{2m}$ and $\Mat{S}(1) = \fourier{\Mat{S}}$~\cite{Lopez19a}. The path $\Mat{S}(t)$ should also have a compatible winding number with the desired overall sign of $\Mat{M}(\fourier{\Mat{S}})$. (Such a path can always be found since the symplectic group is topologically connected.) We then discretize $\Mat{S}(t)$ to obtain $K$ near-identity symplectic matrices as the single-step iterates, namely,
\begin{equation}
    \Mat{S}_j \doteq 
    \Mat{S}\left( \frac{j}{K} \right) \, 
    \Mat{S}^{-1}\left( \frac{j-1}{K} \right)
    , \quad
    j = 1, \ldots, K
    ,
    \label{eq:NIsymp}
\end{equation}

\noindent where we have assumed a uniform step size $\Delta t \doteq 1/K$ for simplicity. Then, since we can decompose $\fourier{\Mat{S}}$ as
\begin{equation}
    \fourier{\Mat{S}} = \Mat{S}_K \ldots \Mat{S}_1
    ,
\end{equation}

\noindent we can compute $\oper{M}(\fourier{\Mat{S}})$ via the iteration scheme
\begin{equation}
    \oper{M}(\fourier{\Mat{S}})
    =
    \oper{M}(\Mat{S}_K) \ldots \oper{M}(\Mat{S}_1) 
    ,
    \label{eq:ITERMT}
\end{equation}

\noindent where each $\oper{M}(\Mat{S}_j)$ is near-identity. Hence, an arbitrary dMT can be approximately computed via the iterated dNIMT algorithm given as
\begin{equation}
    \Mat{M}(\fourier{\Mat{S}})
    \approx
    \Mat{N}(\Mat{S}_K) \ldots \Mat{N}(\Mat{S}_1)
    \label{eq:ITERdNIMT}
    .
\end{equation}

\subsection{Local and global error convergence}

\begin{figure*}
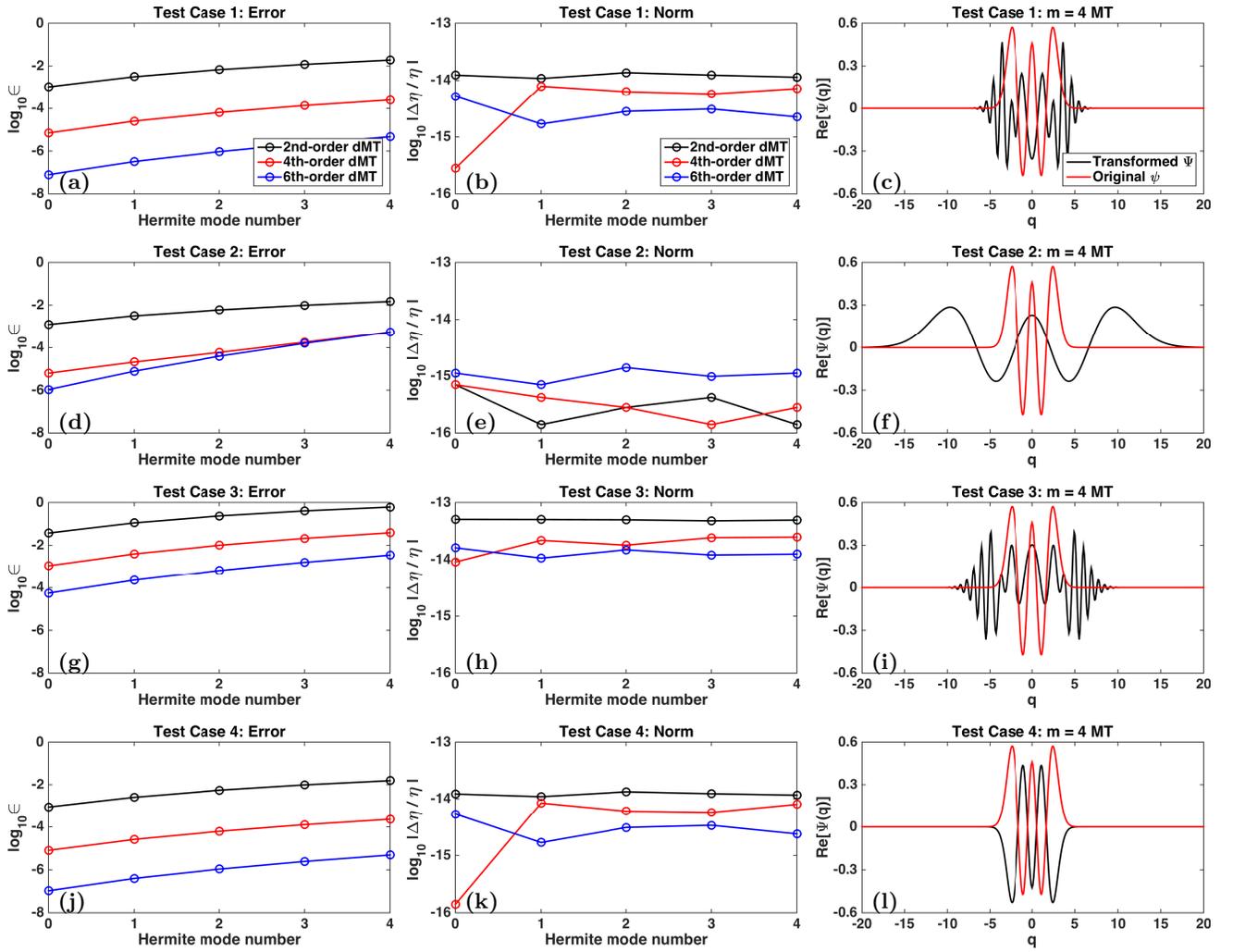

    \begin{overpic}[width=0.32\linewidth,trim={3mm 23mm 16mm 26mm},clip]{errorScan_test1.eps}
        \put(15,10){\textbf{\small(a)}}
    \end{overpic}
    \begin{overpic}[width=0.32\linewidth,trim={3mm 23mm 16mm 26mm},clip]{normScan_test1.eps}
        \put(15,10){\textbf{\small(b)}}
    \end{overpic}
    \begin{overpic}[width=0.32\linewidth,trim={3mm 23mm 16mm 26mm},clip]{compare_m4_test1.eps}
        \put(15,10){\textbf{\small(c)}}
    \end{overpic}
    
    \vspace{2mm}
    \begin{overpic}[width=0.32\linewidth,trim={3mm 23mm 16mm 26mm},clip]{errorScan_test2.eps}
        \put(15,10){\textbf{\small(d)}}
    \end{overpic}
    \begin{overpic}[width=0.32\linewidth,trim={3mm 23mm 16mm 26mm},clip]{normScan_test2.eps}
        \put(15,10){\textbf{\small(e)}}
    \end{overpic}
    \begin{overpic}[width=0.32\linewidth,trim={3mm 23mm 16mm 26mm},clip]{compare_m4_test2.eps}
        \put(15,10){\textbf{\small(f)}}
    \end{overpic}
    
    \vspace{2mm}
    \begin{overpic}[width=0.32\linewidth,trim={3mm 23mm 16mm 26mm},clip]{errorScan_test3.eps}
        \put(15,10){\textbf{\small(g)}}
    \end{overpic}
    \begin{overpic}[width=0.32\linewidth,trim={3mm 23mm 16mm 26mm},clip]{normScan_test3.eps}
        \put(15,10){\textbf{\small(h)}}
    \end{overpic}
    \begin{overpic}[width=0.32\linewidth,trim={3mm 23mm 16mm 26mm},clip]{compare_m4_test3.eps}
        \put(15,10){\textbf{\small(i)}}
    \end{overpic}
    
    \vspace{2mm}
    \begin{overpic}[width=0.32\linewidth,trim={3mm 23mm 16mm 26mm},clip]{errorScan_test4.eps}
        \put(15,10){\textbf{\small(j)}}
    \end{overpic}
    \begin{overpic}[width=0.32\linewidth,trim={3mm 23mm 16mm 26mm},clip]{normScan_test4.eps}
        \put(15,10){\textbf{\small(k)}}
    \end{overpic}
    \begin{overpic}[width=0.32\linewidth,trim={3mm 23mm 16mm 26mm},clip]{compare_m4_test4.eps}
        \put(15,10){\textbf{\small(l)}}
    \end{overpic}
    \caption{\textbf{(a)} Comparing the error $\epsilon$ [\Eq{eq:err}] of the 2nd-order, 4th-order, and 6th-order dMTs applied to the first five HG modes [\Eq{eq:Herm}] for the first test case \eq{eq:test1}. \textbf{(b)} Same as (a), but for the change in norm $\eta$ [\Eq{eq:norm}] rather than $\epsilon$. \textbf{(c)} Comparing the initial and transformed field for the fifth HG mode computed using the 6th-order dMT. For all cases, $q$ is uniformly discretized on the interval $[-20, 20]$ with a step size of $0.1$, so $N = 401$. \textbf{(d) - (f)} Same as (a) - (c), but for the second test case \eq{eq:test2}. \textbf{(g) - (i)} Same as (a) - (c), but for the third test case \eq{eq:test3}. \textbf{(j) - (l)} Same as (a) - (c), but for the fourth test case \eq{eq:test4}.}
    \label{fig:DMT}
\end{figure*}

Analogous to the analysis of finite-difference approximations~\cite{Suli03}, we can consider the local and global convergence of the iterated dNIMT to the dMT. The local error convergence is determined by the truncation error of the Pad\'e approximation. Indeed, since
\begin{align}
    \exp(\Mat{H})
    -
    \left( \IMat{N} - \frac{\Mat{H}}{2} \right)^{-1}
    \left(\IMat{N} + \frac{\Mat{H}}{2}\right)
    &= O(\Delta t^3)
\end{align}

\noindent for any infinitesimal matrix $\Mat{H}$ that satisfies $\| \Mat{H} \| \sim O(\Delta t)$, it is clear that
\begin{equation}
    \Mat{M}(\Mat{S}) = \Mat{N}(\Mat{S}) + O(\Delta t^3)
    \label{eq:localCONVERG}
    .
\end{equation}

\noindent Consequently, the local error between the dNIMT and the dMT converges with a rate of $3$.

To assess the global convergence, let us introduce the sequence of single-step iterates $\{\Mat{m}_j\}$ that are computed during the iterated dNIMT [\Eq{eq:ITERdNIMT}] as
\begin{equation}
    \Mat{m}_{j} = \Mat{N}(\Mat{S}_j) \Mat{m}_{j-1}
    , \quad
    \Mat{m}_0 \doteq \Mat{M}[\Mat{S}(0)]
    ,
    \label{eq:1STEP}
\end{equation}

\noindent along with the sequence $\{\Mat{M}_j \}$ obtained by the iterating the dMT as
\begin{equation}
    \Mat{M}_j = \Mat{M}(\Mat{S}_j) \Mat{M}_{j - 1}
    , \quad
    \Mat{M}_0 = \Mat{m}_0
    .
\end{equation}

\noindent The local convergence of the dNIMT [\Eq{eq:localCONVERG}] implies
\begin{equation}
    \Mat{M}_j 
    = \Mat{N}(\Mat{S}_j) \Mat{M}_{j-1}
    + O(\Delta t ^3)
    .
    \label{eq:1STEPexact}
\end{equation}

\noindent Hence, subtracting \Eq{eq:1STEP} from \Eq{eq:1STEPexact} yields
\begin{equation}
    \Mat{M}_j 
    - \Mat{m}_j = 
    \Mat{N}(\Mat{S}_j )
    \left\{
        \Mat{M}_{j-1}
        - \Mat{m}_{j-1}
        \nullFrac
    \right\}
    + O(\Delta t^3)
    .
\end{equation}

\noindent Since $\Mat{N}(\Mat{S}_j)$ is unitary, we can bound the global error as
\begin{equation}
    \left\| 
        \Mat{M}_j 
        - \Mat{m}_j 
    \right\|
    \le
    \left\| 
        \Mat{M}_{j-1}
        - \Mat{m}_{j-1} 
    \right\|
    + T \Delta t^3
\end{equation}

\noindent for some positive constant $T$. Since $\Mat{m}_0 = \Mat{M}_0$, it follows by induction that
\begin{equation}
    \left\| 
        \Mat{M}_j 
        - \Mat{m}_j 
    \right\| 
    \le j T \Delta t^3
    .
    \label{eq:bound}
\end{equation}

\noindent Finally, for a total number of iterations $K = 1/\Delta t$, \Eq{eq:bound} implies that
\begin{equation}
    \Mat{M}(\Mat{S}_K) \ldots \Mat{M}(\Mat{S}_1)
    = \Mat{N}(\Mat{S}_K) \ldots \Mat{N}(\Mat{S}_1)
    + O(\Delta t^2)
    .
\end{equation}

\noindent Hence, the iterated dNIMT converges to the iterated dMT at a rate of $2$.

The product $\Mat{M}(\Mat{S}_K) \ldots \Mat{M}(\Mat{S}_1)$ may not quite equal $\Mat{M}(\fourier{\Mat{S}})$ because the composition property \eq{eq:MTcomp} might not extend from the MT to the dMT exactly. This is the well-known `loss of additivity' that many discrete MTs experience~\cite{Zhao15}; in our case, it arises from the fact that \Eq{eq:qpCOM} is impossible to satisfy with finite-dimensional representations for $\oper{q}$ and $\oper{p}$~\cite{Weyl50} (although certain identities involving $[\oper{q},\oper{p}]$ can be satisfied with properly chosen finite-difference matrices~\cite{Ninno18}). Still, our numerical results (\Sec{sec:example}) show that the difference between $\Mat{M}_K$ and $\Mat{M}(\fourier{\Mat{S}})$ is negligibly small in many cases, which suggests that the dMT may indeed be additive as well as unitary.


\section{Numerical examples}
\label{sec:example}

Here we demonstrate the performance of the dMT in four examples. Specifically, we use the dMT to compute the transformation of a Hermite--Gauss (HG) laser mode, with transverse field profile given as
\begin{equation}
    \psi_m(q) = 
    \frac{H_m(q)}{\sqrt{2^m m! \sqrt{\pi}}} 
    \exp\left(
        -\frac{q^2}{2}
    \right)
    \label{eq:Herm}
\end{equation}

\noindent (where $H_m$ is the $m$th Hermite polynomial~\cite{Olver10a}) through the paraxial setups corresponding to the following symplectic matrices:
\begin{subequations}
    \begin{align}
        \label{eq:test1}
        \Mat{S}_1
        &=
        \begin{pmatrix}
            1 & 1 \\
            1 & 2
        \end{pmatrix}
        , \\
        \label{eq:test2}
        \Mat{S}_2
        &=
        \begin{pmatrix}
            4 & 0 \\
            0 & 0.25
        \end{pmatrix}
        , \\
        \label{eq:test3}
        \Mat{S}_3
        &=
        \begin{pmatrix}
            0.5 & 2 \\
            -1 & -2
        \end{pmatrix}
        , \\
        \label{eq:test4}
        \Mat{S}_4
        &=
        \frac{1}{\sqrt{2}}
        \begin{pmatrix}
            1 & 1 \\
            -1 & 1
        \end{pmatrix}
        .
    \end{align}
\end{subequations}

\noindent (One such correspondence is discussed following \Eq{eq:decompS}, but others exist as well~\cite{Arsenault80b,Nazarathy82,Liu08,Yasir21}.) For reference, the exact MT for general $\Mat{S}$ of $\psi_m(q)$ given by \Eq{eq:Herm} is
\begin{align}
    &\Psi_m(q) =
    \left(A^2 + B^2\right)^{-1/4}
    \psi_m\left( \frac{q}{\sqrt{A^2 + B^2}} \right)
    \nonumber\\
    &\times
    \exp
    \left[
        i\frac{AC + DB}{2A^2 + 2B^2}q^2
        -i \frac{2m + 1}{2} 
        \tan^{-1}
        \left(
            \frac{B}{A}
        \right)
    \right]
    .
\end{align}

Figure \ref{fig:DMT} shows the error $\epsilon$, defined via the Euclidean $2$-norm as
\begin{equation}
    \epsilon \doteq \frac{\| \Vect{\Psi} - \Vect{\Psi}_\text{exact} \|_2}{\| \Vect{\Psi}_\text{exact} \|_2}
    ,
    \label{eq:err}
\end{equation}

\noindent along with the change in norm $\eta$, defined as 
\begin{equation}
    \Delta \eta \doteq \| \Vect{\Psi} \|_2 - \eta
    , \quad
    \eta \doteq \| \Vect{\psi} \|_2
    ,
    \label{eq:norm}
\end{equation}

\noindent as the dMT for the four test cases is applied to the first five HG modes. Figure \ref{fig:DMT} also shows a comparison between the real parts of $\psi_4(q)$ and $\Psi_4(q)$ for the four test cases. Overall, the norm is preserved to near machine precision, while the error of the dMT decreases as the order of the dMT is increased. The increase in error as $m$ increases is expected since the length scale of $\psi_m(q)$ decreases with $m$, so the finite-difference error at fixed step size consequently increases with $m$. We should note that in these examples, the matrix exponentials are computed using MATLAB's built-in \texttt{expm} method, which uses the standard Pad\'e-based `scaling-squaring' algorithm~\cite{AlMohy09}.

\begin{figure}
    \centering
    \includegraphics[width=\linewidth,trim={2mm 20mm 16mm 26mm},clip]{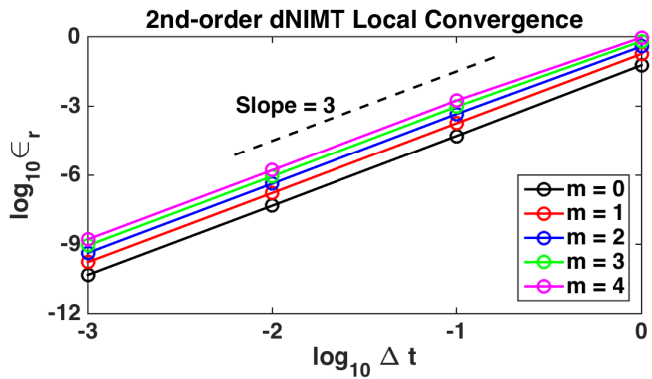}
    \caption{Local (single-step) error convergence of the 2nd-order dNIMT to the dMT for the first five HG modes. The expected convergence rate of $3$ is clearly observed. Although not shown, the 4th-order and the 6th-order dNIMT also exhibit the expected converge rate and in fact, the values of $\epsilon_r$ [\Eq{eq:relERR}] are nearly identical to those of the 2nd-order dNIMT.}
    \label{fig:localNIMT}
\end{figure}

\begin{figure}
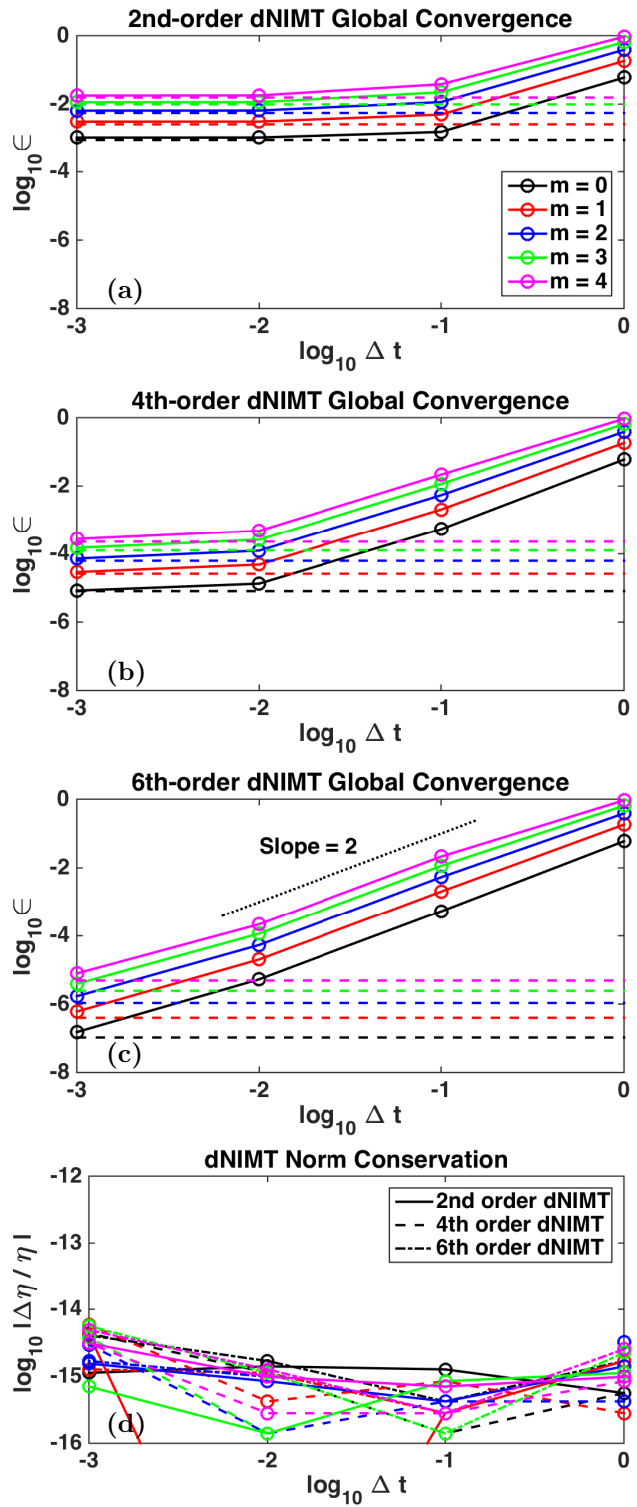

    \begin{overpic}[width=0.95\linewidth,trim={8mm 21mm 17mm 26mm},clip]{globalMAT2converge.eps}
        \put(15,12){\textbf{\normalsize(a)}}
    \end{overpic}
    
    \vspace{2mm}
    \begin{overpic}[width=0.95\linewidth,trim={8mm 21mm 17mm 26mm},clip]{globalMAT4converge.eps}
        \put(15,12){\textbf{\normalsize(b)}}
    \end{overpic}
    
    \vspace{2mm}
    \begin{overpic}[width=0.95\linewidth,trim={8mm 21mm 17mm 26mm},clip]{globalMAT6converge.eps}
        \put(15,12){\textbf{\normalsize(c)}}
    \end{overpic}
    
    \vspace{2mm}
    \begin{overpic}[width=0.95\linewidth,trim={4mm 21mm 17mm 26mm},clip]{globalNORMconserv.eps}
        \put(15,12){\textbf{\normalsize(d)}}
    \end{overpic}
    \caption{\textbf{(a) - (c)} Global convergence of the 2nd-order, 4th-order, and 6th-order dNIMT to the respective dMT for the first five HG modes. The dashed line shows the intrinsic error of the corresponding dMT. As can be seen, the convergence rate achieves the expected value of $2$ before asymptoting to match the error of the dMT. \textbf{(d)} Global norm conservation of the 2nd-order, 4th-order, and 6th-order dNIMT for the first five HG modes. In all cases, the norm is conserved to near machine precision.}
    \label{fig:globalNIMT}
\end{figure}

We next demonstrate the convergence of the iterated dNIMT to the dMT by computing the fourth example \eq{eq:test4} via the path
\begin{equation}
    \Mat{S}(t) = 
    \frac{1}{\sqrt{2}}
    \begin{pmatrix}
        \sqrt{2} + (1 - \sqrt{2}) t & t \\[1mm]
        - t & \frac{2 - t^2}{\sqrt{2} + (1 - \sqrt{2})t}
    \end{pmatrix}
    .
\end{equation}

\noindent One can verify that $\Mat{S}(t)$ is symplectic for all $t \in [0,1]$. Figure~\ref{fig:localNIMT} shows the local error convergence of the dNIMT to the dMT, where the relative error $\epsilon_r$ is defined as
\begin{equation}
    \epsilon_r \doteq \frac{\| \Vect{\Psi}_\text{dNIMT} - \Vect{\Psi}_\text{dMT} \|_2}{\| \Vect{\Psi}_\text{dMT} \|_2}
    \label{eq:relERR}
    .
\end{equation}

\noindent Clearly, the dNIMT converges to the dMT with a rate of $3$, as expected. Note that $\epsilon_r$ is computed for a single iteration, which means that different values of $\Delta t$ correspond to different final transforms $\Mat{S}(\Delta t)$.

Figure~\ref{fig:globalNIMT} shows the global convergence of the iterated dNIMT to the dMT when computing the final transformation $\Mat{S}(1)$ with the specified step size, along with the norm conservation. The asymptote in the convergence plots marks the intrinsic error between the dMT and the exact result, which arises from the use of finite-difference matrices to perform the spatial derivatives. As expected, the iterated dNIMT converges to the dMT at a rate of $2$ and conserves the norm to near machine precision.


\section{Conclusion}
\label{sec:concl}

In this work, we develop a discrete MT that (i) is exactly unitary and (ii) can be approximately computed in time that scales linearly with the number of grid points $N$. This is faster than other known algorithms for the discrete MT, which scale as $O(N \log_2 N)$ due to their similarity with the fast Fourier transform. By using a diagonal Pad\'e approximation for the matrix exponential, we then develop a near-identity approximation to the MT that also (i) is exactly unitary and (ii) can be computed in linear time. This formulation of the NIMT is a marked improvement over that of \Ref{Lopez19a}, which was not unitary and suffered from numerical instability as a consequence. Moreover, we prove that our discrete NIMT always converges to the discrete MT with a second-order accuracy when iterated along a suitable `trajectory' of near-identity symplectic matrices; hence the iterated NIMT can be used to perform finite MTs in $O(KN)$ operations without experiencing any numerical instability, where $K$ is the number of iterations. In particular, the FT can be computed with only two iterations in principle (two iterations ensure each iterate has $\det \textsf{A} \neq 0$), but $K \gg 1$ would be necessary to achieve a reasonable accuracy. We expect these results to be useful for reduced numerical modeling of wave caustics via metaplectic geometrical optics (the general algorithm for which was outlined in \Refs{Lopez20a,Lopez21a}), among other possible applications.

\section*{Funding}

U.S.~DOE Contract No.~DE-AC02-09CH11466.

\section*{Acknowledgements}

The authors thank Laura Xin Zhang for insightful conversations.

\section*{Disclosures}

The authors declare no conflicts of interest.


\appendix

\section{Verification of the operator MT}
\label{app:verify}

We can verify that \Eq{eq:operMT} is indeed a representation of the MT by verifying the following equalities~\cite{Lopez19a}:
\begin{subequations}
    \begin{align}
        \label{eq:Q}
        \oper{M}^\dagger(\Mat{S}) \VectOp{q} \oper{M}(\Mat{S})
        &= \Mat{A} \VectOp{q} + \Mat{B} \VectOp{p}
        , \\
        \label{eq:P}
        \oper{M}^\dagger(\Mat{S}) \VectOp{p} \oper{M}(\Mat{S}) 
        &= \Mat{C} \VectOp{q} + \Mat{D} \VectOp{p}
        .
    \end{align}
\end{subequations}

Let us begin with \Eq{eq:Q}. First, using \Eq{eq:qpCOM}, we can compute the commutator
\begin{equation}
    \left[
        \frac{
            \VectOp{q}^\intercal \left(\log \Mat{A}^{-\intercal}\right) \VectOp{p} 
            + \VectOp{p}^\intercal \left(\log \Mat{A}^{-1}\right) \VectOp{q}
        }{
            2 i
        }
        ,
        \VectOp{q}
    \right]
    =
    (\log \Mat{A})
    \,
    \VectOp{q}
    .
    \label{eq:squeezeQ}
\end{equation}

\noindent Hence, from the Baker--Campbell--Hausdorff (BCH) formula~\cite{Scully12}, it follows by induction that
\begin{equation}
    \oper{D}_\Mat{A}^\dagger \,
    \VectOp{q} \,
    \oper{D}_\Mat{A}
    = 
    \exp\left(\log \Mat{A} \right) \VectOp{q}
    =
    \Mat{A} \VectOp{q}
    ,
    \label{eq:Q1}
\end{equation}

\noindent where we used \Eq{eq:dilOPER} for $\oper{D}_\Mat{A}$. Since functions of $\VectOp{q}$ commute, we trivially obtain
\begin{equation}
    \exp\left(
        - \frac{i}{2} \VectOp{q}^\intercal \Mat{A}^\intercal \Mat{C} \VectOp{q}
    \right)
    \Mat{A} \VectOp{q} 
    \exp\left(
        \frac{i}{2} \VectOp{q}^\intercal \Mat{A}^\intercal \Mat{C} \VectOp{q}
    \right)
    =
    \Mat{A} \VectOp{q} 
    .
    \label{eq:Q2}
\end{equation}

\noindent Next, using \Eq{eq:qpCOM} and the fact that $\Mat{A}^{-1} \Mat{B}$ is symmetric [\Eq{eq:symplec3}], we compute the commutator
\begin{equation}
    \left[
        \frac{i}{2}
        \VectOp{p}^\intercal \Mat{A}^{-1} \Mat{B} \VectOp{p}
        ,
        \Mat{A} \VectOp{q}
        \nullFrac
    \right]
    =
    \Mat{B} \VectOp{p}
    .
    \label{eq:chirpQ}
\end{equation}

\noindent Since the right-hand side does not contain $\VectOp{q}$, the BCH series truncates and we obtain
\begin{align}
    \exp\left(
        \frac{i}{2} \VectOp{p}^\intercal \Mat{A}^{-1} \Mat{B} \VectOp{p}
    \right)
    \Mat{A} \VectOp{q} 
    \exp\left(
        - \frac{i}{2} \VectOp{p}^\intercal \Mat{A}^{-1} \Mat{B} \VectOp{p}
    \right)
    =
    \Mat{A} \VectOp{q} + \Mat{B} \VectOp{p}
    .
    \label{eq:Q3}
\end{align}

\noindent Combining \Eqs{eq:Q1}, \eq{eq:Q2}, and \eq{eq:Q3} yields \Eq{eq:Q}.

Next, let us consider \Eq{eq:P}. Analogous to \Eq{eq:squeezeQ}, we compute the commutator
\begin{equation}
    \left[
        \frac{
            \VectOp{q}^\intercal \left(\log \Mat{A}^{-\intercal}\right) \VectOp{p} 
            + \VectOp{p}^\intercal \left(\log \Mat{A}^{-1}\right) \VectOp{q}
        }{2i}
        ,
        \VectOp{p}
    \right]
    =
    (\log \Mat{A}^{-\intercal})
    \,
    \VectOp{p}
    .
\end{equation}

\noindent By induction, the BCH formula therefore yields
\begin{equation}
    \oper{D}_\Mat{A}^\dagger \,
    \VectOp{p} \,
    \oper{D}_\Mat{A}
    =
    \exp\left(\log \Mat{A}^{-\intercal} \right) \VectOp{p}
    =
    \Mat{A}^{-\intercal} \VectOp{p}
    .
    \label{eq:P1}
\end{equation}

\noindent Analogous to \Eq{eq:chirpQ}, using the fact that $\Mat{A}^\intercal \Mat{C}$ is symmetric [\Eq{eq:symplec5}], we compute the commutator
\begin{equation}
    \left[ 
        -\frac{i}{2} \VectOp{q}^\intercal \Mat{A}^\intercal \Mat{C} \VectOp{q}
        ,
        \Mat{A}^{-\intercal} \VectOp{p}
    \right]
    =
    \Mat{C} \VectOp{q}
    .
\end{equation}

\noindent Since the right-hand side does not contain $\VectOp{p}$, the BCH series truncates and we obtain
\begin{align}
    &\exp\left(
        - \frac{i}{2} \VectOp{q}^\intercal \Mat{A}^{\intercal} \Mat{C} \VectOp{q}
    \right)
    \Mat{A}^{-\intercal} \VectOp{p} 
    \exp\left(
        \frac{i}{2} \VectOp{q}^\intercal \Mat{A}^{\intercal} \Mat{C} \VectOp{q}
    \right)
    \nonumber\\
    &\hspace{5cm}=
    \Mat{A}^{-\intercal} \VectOp{p} + \Mat{C} \VectOp{q}
    .
    \label{eq:P2}
\end{align}

\noindent Similar to \Eq{eq:chirpQ}, we compute the commutator
\begin{equation}
    \com{
        \frac{i}{2}
        \VectOp{p} \Mat{A}^{-1} \Mat{B} \VectOp{p}
    }{
        \Mat{A}^{-\intercal} \VectOp{p} + \Mat{C} \VectOp{q}
    }
    =
    \Mat{C} \Mat{B}^\intercal \Mat{A}^{- \intercal} \VectOp{p}
    .
\end{equation}

\noindent Since the right-hand side does not contain $\VectOp{q}$, the BCH series truncates and we obtain
\begin{align}
    &\exp\left(
        \frac{i}{2} \VectOp{p}^\intercal \Mat{A}^{-1} \Mat{B} \VectOp{p}
    \right)
    \left(
        \Mat{A}^{-\intercal} \VectOp{p} 
        + \Mat{C} \VectOp{q}
    \right)
    \exp\left(
        - \frac{i}{2} \VectOp{p}^\intercal \Mat{A}^{-1} \Mat{B} \VectOp{p}
    \right)
    \nonumber\\
    &\hspace{40mm}=
    \Mat{C} \VectOp{q} 
    + 
    \left(
        \IMat{m} 
        + \Mat{C}\Mat{B}^\intercal 
    \right)\Mat{A}^{-\intercal} \VectOp{p}
    \nonumber\\
    &\hspace{40mm}=
    \Mat{C} \VectOp{q} 
    + \Mat{D} \VectOp{p}
    ,
    \label{eq:P3}
\end{align}

\noindent after using \Eq{eq:symplec1}. Combining \Eqs{eq:P1}, \eq{eq:P2}, and \eq{eq:P3} yields \Eq{eq:P}.

\bibliography{Biblio.bib}

\begin{thebibliography}{43}%
\makeatletter
\providecommand \@ifxundefined [1]{%
 \@ifx{#1\undefined}
}%
\providecommand \@ifnum [1]{%
 \ifnum #1\expandafter \@firstoftwo
 \else \expandafter \@secondoftwo
 \fi
}%
\providecommand \@ifx [1]{%
 \ifx #1\expandafter \@firstoftwo
 \else \expandafter \@secondoftwo
 \fi
}%
\providecommand \natexlab [1]{#1}%
\providecommand \enquote  [1]{``#1''}%
\providecommand \bibnamefont  [1]{#1}%
\providecommand \bibfnamefont [1]{#1}%
\providecommand \citenamefont [1]{#1}%
\providecommand \href@noop [0]{\@secondoftwo}%
\providecommand \href [0]{\begingroup \@sanitize@url \@href}%
\providecommand \@href[1]{\@@startlink{#1}\@@href}%
\providecommand \@@href[1]{\endgroup#1\@@endlink}%
\providecommand \@sanitize@url [0]{\catcode `\\12\catcode `\$12\catcode
  `\&12\catcode `\#12\catcode `\^12\catcode `\_12\catcode `\%12\relax}%
\providecommand \@@startlink[1]{}%
\providecommand \@@endlink[0]{}%
\providecommand \url  [0]{\begingroup\@sanitize@url \@url }%
\providecommand \@url [1]{\endgroup\@href {#1}{\urlprefix }}%
\providecommand \urlprefix  [0]{URL }%
\providecommand \Eprint [0]{\href }%
\providecommand \doibase [0]{http://dx.doi.org/}%
\providecommand \selectlanguage [0]{\@gobble}%
\providecommand \bibinfo  [0]{\@secondoftwo}%
\providecommand \bibfield  [0]{\@secondoftwo}%
\providecommand \translation [1]{[#1]}%
\providecommand \BibitemOpen [0]{}%
\providecommand \bibitemStop [0]{}%
\providecommand \bibitemNoStop [0]{.\EOS\space}%
\providecommand \EOS [0]{\spacefactor3000\relax}%
\providecommand \BibitemShut  [1]{\csname bibitem#1\endcsname}%
\let\auto@bib@innerbib\@empty
\bibitem [{\citenamefont {Healy}\ \emph {et~al.}(2016)\citenamefont {Healy},
  \citenamefont {Kutay}, \citenamefont {Ozaktas},\ and\ \citenamefont
  {Sheridan}}]{Healy16}%
  \BibitemOpen
  \bibinfo {editor} {\bibfnamefont {J.~J.}\ \bibnamefont {Healy}}, \bibinfo
  {editor} {\bibfnamefont {M.~A.}\ \bibnamefont {Kutay}}, \bibinfo {editor}
  {\bibfnamefont {H.~M.}\ \bibnamefont {Ozaktas}}, \ and\ \bibinfo {editor}
  {\bibfnamefont {J.~T.}\ \bibnamefont {Sheridan}},\ eds.,\ \href {\doibase
  10.1007/978-1-4939-3028-9} {\emph {\bibinfo {title} {Linear Canonical
  Transforms: Theory and Applications}}}\ (\bibinfo  {publisher} {New York:
  Springer},\ \bibinfo {year} {2016})\BibitemShut {NoStop}%
\bibitem [{\citenamefont {Collins}(1970)}]{Collins70}%
  \BibitemOpen
  \bibfield  {author} {\bibinfo {author} {\bibfnamefont {S.~A.}\ \bibnamefont
  {Collins}},\ }\href {\doibase 10.1364/JOSA.60.001168} {\bibfield  {journal}
  {\bibinfo  {journal} {J. Opt. Soc. Am.}\ }\textbf {\bibinfo {volume} {60}},\
  \bibinfo {pages} {1168} (\bibinfo {year} {1970})}\BibitemShut {NoStop}%
\bibitem [{\citenamefont {Bacry}\ and\ \citenamefont
  {Cadilhac}(1981)}]{Bacry81}%
  \BibitemOpen
  \bibfield  {author} {\bibinfo {author} {\bibfnamefont {H.}~\bibnamefont
  {Bacry}}\ and\ \bibinfo {author} {\bibfnamefont {M.}~\bibnamefont
  {Cadilhac}},\ }\href {\doibase 10.1103/PhysRevA.23.2533} {\bibfield
  {journal} {\bibinfo  {journal} {Phys. Rev. A}\ }\textbf {\bibinfo {volume}
  {23}},\ \bibinfo {pages} {2533} (\bibinfo {year} {1981})}\BibitemShut
  {NoStop}%
\bibitem [{\citenamefont {Simon}\ and\ \citenamefont {Wolf}(2000)}]{Simon00}%
  \BibitemOpen
  \bibfield  {author} {\bibinfo {author} {\bibfnamefont {R.}~\bibnamefont
  {Simon}}\ and\ \bibinfo {author} {\bibfnamefont {K.~B.}\ \bibnamefont
  {Wolf}},\ }\href {\doibase 10.1364/JOSAA.17.000342} {\bibfield  {journal}
  {\bibinfo  {journal} {J. Opt. Soc. Am. A}\ }\textbf {\bibinfo {volume}
  {17}},\ \bibinfo {pages} {342} (\bibinfo {year} {2000})}\BibitemShut
  {NoStop}%
\bibitem [{\citenamefont {Wolf}(2018)}]{Wolf18}%
  \BibitemOpen
  \bibfield  {author} {\bibinfo {author} {\bibfnamefont {K.~B.}\ \bibnamefont
  {Wolf}},\ }in\ \href {\doibase 10.1016/B978-0-12-803581-8.09380-2} {\emph
  {\bibinfo {booktitle} {Encyclopedia of Modern Optics}}},\ Vol.~\bibinfo
  {volume} {4}\ (\bibinfo  {publisher} {Oxford: Elsevier},\ \bibinfo {year}
  {2018})\ \bibinfo {edition} {2nd}\ ed.,\ p.\ \bibinfo {pages}
  {199}\BibitemShut {NoStop}%
\bibitem [{\citenamefont {Lopez}\ and\ \citenamefont {Dodin}(2020)}]{Lopez20a}%
  \BibitemOpen
  \bibfield  {author} {\bibinfo {author} {\bibfnamefont {N.~A.}\ \bibnamefont
  {Lopez}}\ and\ \bibinfo {author} {\bibfnamefont {I.~Y.}\ \bibnamefont
  {Dodin}},\ }\href {\doibase 10.1088/1367-2630/aba91a} {\bibfield  {journal}
  {\bibinfo  {journal} {New J. Phys.}\ }\textbf {\bibinfo {volume} {22}},\
  \bibinfo {pages} {083078} (\bibinfo {year} {2020})}\BibitemShut {NoStop}%
\bibitem [{\citenamefont {Lopez}\ and\ \citenamefont {Dodin}(2021)}]{Lopez21a}%
  \BibitemOpen
  \bibfield  {author} {\bibinfo {author} {\bibfnamefont {N.~A.}\ \bibnamefont
  {Lopez}}\ and\ \bibinfo {author} {\bibfnamefont {I.~Y.}\ \bibnamefont
  {Dodin}},\ }\href {\doibase 10.1088/2040-8986/abd1ce} {\bibfield  {journal}
  {\bibinfo  {journal} {J. Opt.}\ }\textbf {\bibinfo {volume} {23}},\ \bibinfo
  {pages} {025601} (\bibinfo {year} {2021})}\BibitemShut {NoStop}%
\bibitem [{\citenamefont {Ozaktas}\ \emph {et~al.}(1996)\citenamefont
  {Ozaktas}, \citenamefont {Arikan}, \citenamefont {Kutay},\ and\ \citenamefont
  {Bozdagi}}]{Ozaktas96}%
  \BibitemOpen
  \bibfield  {author} {\bibinfo {author} {\bibfnamefont {H.~M.}\ \bibnamefont
  {Ozaktas}}, \bibinfo {author} {\bibfnamefont {O.}~\bibnamefont {Arikan}},
  \bibinfo {author} {\bibfnamefont {M.~A.}\ \bibnamefont {Kutay}}, \ and\
  \bibinfo {author} {\bibfnamefont {G.}~\bibnamefont {Bozdagi}},\ }\href
  {\doibase 10.1109/78.536672} {\bibfield  {journal} {\bibinfo  {journal} {IEEE
  Trans. Signal Processing}\ }\textbf {\bibinfo {volume} {44}},\ \bibinfo
  {pages} {2141} (\bibinfo {year} {1996})}\BibitemShut {NoStop}%
\bibitem [{\citenamefont {Hennelly}\ and\ \citenamefont
  {Sheridan}(2005)}]{Hennelly05b}%
  \BibitemOpen
  \bibfield  {author} {\bibinfo {author} {\bibfnamefont {B.~M.}\ \bibnamefont
  {Hennelly}}\ and\ \bibinfo {author} {\bibfnamefont {J.~T.}\ \bibnamefont
  {Sheridan}},\ }\href {\doibase 10.1364/JOSAA.22.000928} {\bibfield  {journal}
  {\bibinfo  {journal} {J. Opt. Soc. Am. A}\ }\textbf {\bibinfo {volume}
  {22}},\ \bibinfo {pages} {928} (\bibinfo {year} {2005})}\BibitemShut
  {NoStop}%
\bibitem [{\citenamefont {Healy}\ and\ \citenamefont
  {Sheridan}(2010)}]{Healy10}%
  \BibitemOpen
  \bibfield  {author} {\bibinfo {author} {\bibfnamefont {J.~J.}\ \bibnamefont
  {Healy}}\ and\ \bibinfo {author} {\bibfnamefont {J.~T.}\ \bibnamefont
  {Sheridan}},\ }\href {\doibase 10.1364/JOSAA.27.000021} {\bibfield  {journal}
  {\bibinfo  {journal} {J. Opt. Soc. Am. A}\ }\textbf {\bibinfo {volume}
  {27}},\ \bibinfo {pages} {21} (\bibinfo {year} {2010})}\BibitemShut {NoStop}%
\bibitem [{\citenamefont {Koc}\ \emph {et~al.}(2010)\citenamefont {Koc},
  \citenamefont {Ozaktas},\ and\ \citenamefont {Hesselink}}]{Koc10a}%
  \BibitemOpen
  \bibfield  {author} {\bibinfo {author} {\bibfnamefont {A.}~\bibnamefont
  {Koc}}, \bibinfo {author} {\bibfnamefont {H.~M.}\ \bibnamefont {Ozaktas}}, \
  and\ \bibinfo {author} {\bibfnamefont {L.}~\bibnamefont {Hesselink}},\ }\href
  {\doibase 10.1364/JOSAA.27.001288} {\bibfield  {journal} {\bibinfo  {journal}
  {J. Opt. Soc. Am. A}\ }\textbf {\bibinfo {volume} {27}},\ \bibinfo {pages}
  {1288} (\bibinfo {year} {2010})}\BibitemShut {NoStop}%
\bibitem [{\citenamefont {Ding}\ \emph {et~al.}(2012)\citenamefont {Ding},
  \citenamefont {Pei},\ and\ \citenamefont {Liu}}]{Ding12}%
  \BibitemOpen
  \bibfield  {author} {\bibinfo {author} {\bibfnamefont {J.-J.}\ \bibnamefont
  {Ding}}, \bibinfo {author} {\bibfnamefont {S.-C.}\ \bibnamefont {Pei}}, \
  and\ \bibinfo {author} {\bibfnamefont {C.-L.}\ \bibnamefont {Liu}},\ }\href
  {\doibase 10.1364/JOSAA.29.001615} {\bibfield  {journal} {\bibinfo  {journal}
  {J. Opt. Soc. Am. A}\ }\textbf {\bibinfo {volume} {29}},\ \bibinfo {pages}
  {1615} (\bibinfo {year} {2012})}\BibitemShut {NoStop}%
\bibitem [{\citenamefont {Pei}\ and\ \citenamefont {Huang}(2016)}]{Pei16}%
  \BibitemOpen
  \bibfield  {author} {\bibinfo {author} {\bibfnamefont {S.-C.}\ \bibnamefont
  {Pei}}\ and\ \bibinfo {author} {\bibfnamefont {S.-G.}\ \bibnamefont
  {Huang}},\ }\href {\doibase 10.1364/JOSAA.33.000214} {\bibfield  {journal}
  {\bibinfo  {journal} {J. Opt. Soc. Am. A}\ }\textbf {\bibinfo {volume}
  {33}},\ \bibinfo {pages} {214} (\bibinfo {year} {2016})}\BibitemShut
  {NoStop}%
\bibitem [{\citenamefont {Sun}\ and\ \citenamefont {Li}(2018)}]{Sun18a}%
  \BibitemOpen
  \bibfield  {author} {\bibinfo {author} {\bibfnamefont {Y.-N.}\ \bibnamefont
  {Sun}}\ and\ \bibinfo {author} {\bibfnamefont {B.-Z.}\ \bibnamefont {Li}},\
  }\href {\doibase 10.1364/JOSAA.35.001346} {\bibfield  {journal} {\bibinfo
  {journal} {J. Opt. Soc. Am. A}\ }\textbf {\bibinfo {volume} {35}},\ \bibinfo
  {pages} {1346} (\bibinfo {year} {2018})}\BibitemShut {NoStop}%
\bibitem [{\citenamefont {Healy}(2018)}]{Healy18}%
  \BibitemOpen
  \bibfield  {author} {\bibinfo {author} {\bibfnamefont {J.~J.}\ \bibnamefont
  {Healy}},\ }\href {\doibase 10.1088/2040-8986/aa9e20} {\bibfield  {journal}
  {\bibinfo  {journal} {J. Opt.}\ }\textbf {\bibinfo {volume} {20}},\ \bibinfo
  {pages} {014008} (\bibinfo {year} {2018})}\BibitemShut {NoStop}%
\bibitem [{\citenamefont {Lopez}\ and\ \citenamefont {Dodin}(2019)}]{Lopez19a}%
  \BibitemOpen
  \bibfield  {author} {\bibinfo {author} {\bibfnamefont {N.~A.}\ \bibnamefont
  {Lopez}}\ and\ \bibinfo {author} {\bibfnamefont {I.~Y.}\ \bibnamefont
  {Dodin}},\ }\href {\doibase 10.1364/JOSAA.36.001846} {\bibfield  {journal}
  {\bibinfo  {journal} {J. Opt. Soc. Am. A}\ }\textbf {\bibinfo {volume}
  {36}},\ \bibinfo {pages} {1846} (\bibinfo {year} {2019})}\BibitemShut
  {NoStop}%
\bibitem [{\citenamefont {Kogelnik}\ and\ \citenamefont
  {Li}(1966)}]{Kogelnik66}%
  \BibitemOpen
  \bibfield  {author} {\bibinfo {author} {\bibfnamefont {H.}~\bibnamefont
  {Kogelnik}}\ and\ \bibinfo {author} {\bibfnamefont {T.}~\bibnamefont {Li}},\
  }\href {\doibase 10.1364/AO.5.001550} {\bibfield  {journal} {\bibinfo
  {journal} {Appl. Opt.}\ }\textbf {\bibinfo {volume} {5}},\ \bibinfo {pages}
  {1550} (\bibinfo {year} {1966})}\BibitemShut {NoStop}%
\bibitem [{\citenamefont {Luneburg}(1964)}]{Luneburg64}%
  \BibitemOpen
  \bibfield  {author} {\bibinfo {author} {\bibfnamefont {R.~K.}\ \bibnamefont
  {Luneburg}},\ }\href@noop {} {\emph {\bibinfo {title} {Mathematical Theory of
  Optics}}}\ (\bibinfo  {publisher} {Berkeley: U. California Press},\ \bibinfo
  {year} {1964})\BibitemShut {NoStop}%
\bibitem [{\citenamefont {Moshinsky}\ and\ \citenamefont
  {Quesne}(1971)}]{Moshinsky71}%
  \BibitemOpen
  \bibfield  {author} {\bibinfo {author} {\bibfnamefont {M.}~\bibnamefont
  {Moshinsky}}\ and\ \bibinfo {author} {\bibfnamefont {C.}~\bibnamefont
  {Quesne}},\ }\href {\doibase 10.1063/1.1665805} {\bibfield  {journal}
  {\bibinfo  {journal} {J. Math. Phys.}\ }\textbf {\bibinfo {volume} {12}},\
  \bibinfo {pages} {1772} (\bibinfo {year} {1971})}\BibitemShut {NoStop}%
\bibitem [{\citenamefont {Littlejohn}(1986)}]{Littlejohn86a}%
  \BibitemOpen
  \bibfield  {author} {\bibinfo {author} {\bibfnamefont {R.~G.}\ \bibnamefont
  {Littlejohn}},\ }\href {\doibase 10.1016/0370-1573(86)90103-1} {\bibfield
  {journal} {\bibinfo  {journal} {Phys. Rep.}\ }\textbf {\bibinfo {volume}
  {138}},\ \bibinfo {pages} {193} (\bibinfo {year} {1986})}\BibitemShut
  {NoStop}%
\bibitem [{\citenamefont {Stoler}(1981)}]{Stoler81}%
  \BibitemOpen
  \bibfield  {author} {\bibinfo {author} {\bibfnamefont {D.}~\bibnamefont
  {Stoler}},\ }\href {\doibase 10.1364/JOSA.71.000334} {\bibfield  {journal}
  {\bibinfo  {journal} {J. Opt. Soc. Am.}\ }\textbf {\bibinfo {volume} {71}},\
  \bibinfo {pages} {334} (\bibinfo {year} {1981})}\BibitemShut {NoStop}%
\bibitem [{\citenamefont {Scully}\ and\ \citenamefont
  {Zubairy}(2012)}]{Scully12}%
  \BibitemOpen
  \bibfield  {author} {\bibinfo {author} {\bibfnamefont {M.~O.}\ \bibnamefont
  {Scully}}\ and\ \bibinfo {author} {\bibfnamefont {M.~S.}\ \bibnamefont
  {Zubairy}},\ }\href {\doibase 10.1017/CBO9780511813993} {\emph {\bibinfo
  {title} {Quantum Optics}}}\ (\bibinfo  {publisher} {Cambridge: Cambridge
  University Press},\ \bibinfo {year} {2012})\BibitemShut {NoStop}%
\bibitem [{\citenamefont {Strang}\ and\ \citenamefont
  {MacNamara}(2014)}]{Strang14}%
  \BibitemOpen
  \bibfield  {author} {\bibinfo {author} {\bibfnamefont {G.}~\bibnamefont
  {Strang}}\ and\ \bibinfo {author} {\bibfnamefont {S.}~\bibnamefont
  {MacNamara}},\ }\href {\doibase 10.1137/120897572} {\bibfield  {journal}
  {\bibinfo  {journal} {SIAM Rev.}\ }\textbf {\bibinfo {volume} {56}},\
  \bibinfo {pages} {525} (\bibinfo {year} {2014})}\BibitemShut {NoStop}%
\bibitem [{\citenamefont {Iserles}(2000)}]{Iserles00}%
  \BibitemOpen
  \bibfield  {author} {\bibinfo {author} {\bibfnamefont {A.}~\bibnamefont
  {Iserles}},\ }\href@noop {} {\bibfield  {journal} {\bibinfo  {journal} {New
  Zealand J. Math.}\ }\textbf {\bibinfo {volume} {29}},\ \bibinfo {pages} {177}
  (\bibinfo {year} {2000})}\BibitemShut {NoStop}%
\bibitem [{\citenamefont {Benzi}\ and\ \citenamefont {Razouk}(2007)}]{Benzi07}%
  \BibitemOpen
  \bibfield  {author} {\bibinfo {author} {\bibfnamefont {M.}~\bibnamefont
  {Benzi}}\ and\ \bibinfo {author} {\bibfnamefont {N.}~\bibnamefont {Razouk}},\
  }\href@noop {} {\bibfield  {journal} {\bibinfo  {journal} {Electron. Trans.
  Numer. Anal.}\ }\textbf {\bibinfo {volume} {28}},\ \bibinfo {pages} {16}
  (\bibinfo {year} {2007})}\BibitemShut {NoStop}%
\bibitem [{\citenamefont {{Al-Mohy}}\ and\ \citenamefont
  {Higham}(2009)}]{AlMohy09}%
  \BibitemOpen
  \bibfield  {author} {\bibinfo {author} {\bibfnamefont {A.~H.}\ \bibnamefont
  {{Al-Mohy}}}\ and\ \bibinfo {author} {\bibfnamefont {N.~J.}\ \bibnamefont
  {Higham}},\ }\href {\doibase 10.1137/09074721X} {\bibfield  {journal}
  {\bibinfo  {journal} {SIAM J. Matrix Anal. Appl.}\ }\textbf {\bibinfo
  {volume} {31}},\ \bibinfo {pages} {970} (\bibinfo {year} {2009})}\BibitemShut
  {NoStop}%
\bibitem [{\citenamefont {{Al-Mohy}}\ and\ \citenamefont
  {Higham}(2011)}]{AlMohy11}%
  \BibitemOpen
  \bibfield  {author} {\bibinfo {author} {\bibfnamefont {A.~H.}\ \bibnamefont
  {{Al-Mohy}}}\ and\ \bibinfo {author} {\bibfnamefont {N.~J.}\ \bibnamefont
  {Higham}},\ }\href {\doibase 10.1137/100788860} {\bibfield  {journal}
  {\bibinfo  {journal} {SIAM J. Sci. Comput.}\ }\textbf {\bibinfo {volume}
  {33}},\ \bibinfo {pages} {488} (\bibinfo {year} {2011})}\BibitemShut
  {NoStop}%
\bibitem [{\citenamefont {Press}\ \emph {et~al.}(2007)\citenamefont {Press},
  \citenamefont {Teukolsky}, \citenamefont {Vetterling},\ and\ \citenamefont
  {Flannery}}]{Press07}%
  \BibitemOpen
  \bibfield  {author} {\bibinfo {author} {\bibfnamefont {W.~H.}\ \bibnamefont
  {Press}}, \bibinfo {author} {\bibfnamefont {S.~A.}\ \bibnamefont
  {Teukolsky}}, \bibinfo {author} {\bibfnamefont {W.~T.}\ \bibnamefont
  {Vetterling}}, \ and\ \bibinfo {author} {\bibfnamefont {B.~P.}\ \bibnamefont
  {Flannery}},\ }\href@noop {} {\emph {\bibinfo {title} {Numerical Recipes}}},\
  \bibinfo {edition} {3rd}\ ed.\ (\bibinfo  {publisher} {Cambridge: Cambridge
  University Press},\ \bibinfo {year} {2007})\BibitemShut {NoStop}%
\bibitem [{\citenamefont {Olver}\ \emph {et~al.}(2010)\citenamefont {Olver},
  \citenamefont {Lozier}, \citenamefont {Boisvert},\ and\ \citenamefont
  {Clark}}]{Olver10a}%
  \BibitemOpen
  \bibfield  {author} {\bibinfo {author} {\bibfnamefont {F.~W.~J.}\
  \bibnamefont {Olver}}, \bibinfo {author} {\bibfnamefont {D.~W.}\ \bibnamefont
  {Lozier}}, \bibinfo {author} {\bibfnamefont {R.~F.}\ \bibnamefont
  {Boisvert}}, \ and\ \bibinfo {author} {\bibfnamefont {C.~W.}\ \bibnamefont
  {Clark}},\ }\href@noop {} {\emph {\bibinfo {title} {NIST Handbook of
  Mathematical Functions}}}\ (\bibinfo  {publisher} {Cambridge: Cambridge
  University Press},\ \bibinfo {year} {2010})\BibitemShut {NoStop}%
\bibitem [{\citenamefont {Eves}(1980)}]{Eves80}%
  \BibitemOpen
  \bibfield  {author} {\bibinfo {author} {\bibfnamefont {H.}~\bibnamefont
  {Eves}},\ }\href@noop {} {\emph {\bibinfo {title} {Elementary Matrix
  Theory}}}\ (\bibinfo  {publisher} {New York: Dover},\ \bibinfo {year}
  {1980})\BibitemShut {NoStop}%
\bibitem [{\citenamefont {Diele}\ \emph {et~al.}(1998)\citenamefont {Diele},
  \citenamefont {Lopez},\ and\ \citenamefont {Peluso}}]{Diele98}%
  \BibitemOpen
  \bibfield  {author} {\bibinfo {author} {\bibfnamefont {F.}~\bibnamefont
  {Diele}}, \bibinfo {author} {\bibfnamefont {L.}~\bibnamefont {Lopez}}, \ and\
  \bibinfo {author} {\bibfnamefont {R.}~\bibnamefont {Peluso}},\ }\href
  {\doibase 10.1023/A:1018908700358} {\bibfield  {journal} {\bibinfo  {journal}
  {Adv. Comp. Math.}\ }\textbf {\bibinfo {volume} {8}},\ \bibinfo {pages} {317}
  (\bibinfo {year} {1998})}\BibitemShut {NoStop}%
\bibitem [{\citenamefont {Iserles}(2001)}]{Iserles01}%
  \BibitemOpen
  \bibfield  {author} {\bibinfo {author} {\bibfnamefont {A.}~\bibnamefont
  {Iserles}},\ }\href {\doibase 10.1007/s102080010003} {\bibfield  {journal}
  {\bibinfo  {journal} {Found. Comput. Math}\ }\textbf {\bibinfo {volume}
  {1}},\ \bibinfo {pages} {129} (\bibinfo {year} {2001})}\BibitemShut {NoStop}%
\bibitem [{\citenamefont {Golub}\ and\ \citenamefont {{Van
  Loan}}(2013)}]{Golub13}%
  \BibitemOpen
  \bibfield  {author} {\bibinfo {author} {\bibfnamefont {G.~H.}\ \bibnamefont
  {Golub}}\ and\ \bibinfo {author} {\bibfnamefont {C.~F.}\ \bibnamefont {{Van
  Loan}}},\ }\href@noop {} {\emph {\bibinfo {title} {Matrix Computations}}},\
  \bibinfo {edition} {4th}\ ed.\ (\bibinfo  {publisher} {Baltimore: Johns
  Hopkins University Press},\ \bibinfo {year} {2013})\BibitemShut {NoStop}%
\bibitem [{\citenamefont {Zhang}\ \emph {et~al.}(2020)\citenamefont {Zhang},
  \citenamefont {Fu},\ and\ \citenamefont {Qin}}]{Zhang20}%
  \BibitemOpen
  \bibfield  {author} {\bibinfo {author} {\bibfnamefont {X.}~\bibnamefont
  {Zhang}}, \bibinfo {author} {\bibfnamefont {Y.}~\bibnamefont {Fu}}, \ and\
  \bibinfo {author} {\bibfnamefont {H.}~\bibnamefont {Qin}},\ }\href {\doibase
  10.1103/PhysRevE.102.033302} {\bibfield  {journal} {\bibinfo  {journal}
  {Phys. Rev. E}\ }\textbf {\bibinfo {volume} {102}},\ \bibinfo {pages}
  {033302} (\bibinfo {year} {2020})}\BibitemShut {NoStop}%
\bibitem [{\citenamefont {Fu}\ \emph {et~al.}(2020)\citenamefont {Fu},
  \citenamefont {Zhang},\ and\ \citenamefont {Qin}}]{Fu20}%
  \BibitemOpen
  \bibfield  {author} {\bibinfo {author} {\bibfnamefont {Y.}~\bibnamefont
  {Fu}}, \bibinfo {author} {\bibfnamefont {X.}~\bibnamefont {Zhang}}, \ and\
  \bibinfo {author} {\bibfnamefont {H.}~\bibnamefont {Qin}},\ }\href@noop {}
  {\bibfield  {journal} {\bibinfo  {journal} {arXiv:2010.12920}\ } (\bibinfo
  {year} {2020})}\BibitemShut {NoStop}%
\bibitem [{\citenamefont {Suli}\ and\ \citenamefont {Mayers}(2003)}]{Suli03}%
  \BibitemOpen
  \bibfield  {author} {\bibinfo {author} {\bibfnamefont {E.}~\bibnamefont
  {Suli}}\ and\ \bibinfo {author} {\bibfnamefont {D.~F.}\ \bibnamefont
  {Mayers}},\ }\href@noop {} {\emph {\bibinfo {title} {An Introduction to
  Numerical Analysis}}}\ (\bibinfo  {publisher} {Cambridge: Cambridge
  University Press},\ \bibinfo {year} {2003})\BibitemShut {NoStop}%
\bibitem [{\citenamefont {Zhao}\ \emph {et~al.}(2015)\citenamefont {Zhao},
  \citenamefont {Healy},\ and\ \citenamefont {Sheridan}}]{Zhao15}%
  \BibitemOpen
  \bibfield  {author} {\bibinfo {author} {\bibfnamefont {L.}~\bibnamefont
  {Zhao}}, \bibinfo {author} {\bibfnamefont {J.~J.}\ \bibnamefont {Healy}}, \
  and\ \bibinfo {author} {\bibfnamefont {J.~T.}\ \bibnamefont {Sheridan}},\
  }\href {\doibase 10.1364/AO.54.009960} {\bibfield  {journal} {\bibinfo
  {journal} {Appl. Opt.}\ }\textbf {\bibinfo {volume} {54}},\ \bibinfo {pages}
  {9960} (\bibinfo {year} {2015})}\BibitemShut {NoStop}%
\bibitem [{\citenamefont {Weyl}(1950)}]{Weyl50}%
  \BibitemOpen
  \bibfield  {author} {\bibinfo {author} {\bibfnamefont {H.}~\bibnamefont
  {Weyl}},\ }\href@noop {} {\emph {\bibinfo {title} {The Theory of Groups and
  Quantum Mechanics}}}\ (\bibinfo  {publisher} {New York: Dover},\ \bibinfo
  {year} {1950})\BibitemShut {NoStop}%
\bibitem [{\citenamefont {Ninno}\ \emph {et~al.}(2018)\citenamefont {Ninno},
  \citenamefont {Cantele},\ and\ \citenamefont {Trani}}]{Ninno18}%
  \BibitemOpen
  \bibfield  {author} {\bibinfo {author} {\bibfnamefont {D.}~\bibnamefont
  {Ninno}}, \bibinfo {author} {\bibfnamefont {G.}~\bibnamefont {Cantele}}, \
  and\ \bibinfo {author} {\bibfnamefont {F.}~\bibnamefont {Trani}},\ }\href
  {\doibase 10.1002/jcc.25208} {\bibfield  {journal} {\bibinfo  {journal} {J.
  Comput. Chem.}\ }\textbf {\bibinfo {volume} {39}},\ \bibinfo {pages} {1406}
  (\bibinfo {year} {2018})}\BibitemShut {NoStop}%
\bibitem [{\citenamefont {Arsenault}(1980)}]{Arsenault80b}%
  \BibitemOpen
  \bibfield  {author} {\bibinfo {author} {\bibfnamefont {H.~H.}\ \bibnamefont
  {Arsenault}},\ }\href {\doibase 10.1119/1.12112} {\bibfield  {journal}
  {\bibinfo  {journal} {Am. J. Phys.}\ }\textbf {\bibinfo {volume} {48}},\
  \bibinfo {pages} {397} (\bibinfo {year} {1980})}\BibitemShut {NoStop}%
\bibitem [{\citenamefont {Nazarathy}\ and\ \citenamefont
  {Shamir}(1982)}]{Nazarathy82}%
  \BibitemOpen
  \bibfield  {author} {\bibinfo {author} {\bibfnamefont {M.}~\bibnamefont
  {Nazarathy}}\ and\ \bibinfo {author} {\bibfnamefont {J.}~\bibnamefont
  {Shamir}},\ }\href {\doibase 10.1364/JOSA.72.000356} {\bibfield  {journal}
  {\bibinfo  {journal} {J. Opt. Soc. Am.}\ }\textbf {\bibinfo {volume} {72}},\
  \bibinfo {pages} {356} (\bibinfo {year} {1982})}\BibitemShut {NoStop}%
\bibitem [{\citenamefont {Liu}\ and\ \citenamefont {Brenner}(2008)}]{Liu08}%
  \BibitemOpen
  \bibfield  {author} {\bibinfo {author} {\bibfnamefont {X.}~\bibnamefont
  {Liu}}\ and\ \bibinfo {author} {\bibfnamefont {K.-H.}\ \bibnamefont
  {Brenner}},\ }\href {\doibase 10.1364/AO.47.000E88} {\bibfield  {journal}
  {\bibinfo  {journal} {Appl. Opt.}\ }\textbf {\bibinfo {volume} {47}},\
  \bibinfo {pages} {E88} (\bibinfo {year} {2008})}\BibitemShut {NoStop}%
\bibitem [{\citenamefont {Yasir}(2021)}]{Yasir21}%
  \BibitemOpen
  \bibfield  {author} {\bibinfo {author} {\bibfnamefont {P.~A.~A.}\
  \bibnamefont {Yasir}},\ }\href {\doibase 10.1364/JOSAA.404552} {\bibfield
  {journal} {\bibinfo  {journal} {J. Opt. Soc. Am. A}\ }\textbf {\bibinfo
  {volume} {38}},\ \bibinfo {pages} {42} (\bibinfo {year} {2021})}\BibitemShut
  {NoStop}%
\end{thebibliography}%
\bibliographystyle{apsrev4-1}
\end{document}